\documentclass[a4paper,twocolumn,10pt,accepted=2022-04-05]{quantumarticle}
\pdfoutput=1
\usepackage[utf8]{inputenc}
\usepackage[english]{babel}
\usepackage[T1]{fontenc}
\usepackage{amsmath}
\usepackage{hyperref}
\usepackage{tikz}
\usepackage{lipsum}
\newtheorem{theorem}{Theorem}

\newtheorem{lemma}[theorem]{Lemma}
\usepackage{makecell}
\usepackage{bbm}
\usepackage{mathrsfs}
\usepackage{amsfonts}
\usepackage{color}
\usepackage{graphicx}
\newcommand{\xv}{\boldsymbol x}

\newcommand{\Dm}{{\boldsymbol{D}}}

\newcommand{\be}{\begin{equation}}
\newcommand{\ee}{\end{equation}}
\newcommand{\ba}{\begin{eqnarray}}
\newcommand{\ea}{\end{eqnarray}}
\newcommand{\ddelta}{\boldsymbol{\delta}}
\newcommand{\RNum}[1]{\uppercase\expandafter{\romannumeral #1\relax}}
\makeatletter
\newdimen\scalemath@axis
\newcommand*{\scalemath}[3]{%
  #1{%
    \mathpalette{\scalemath@aux{#2}}{#3}%
  }%
}
\newcommand*{\scalemath@aux}[3]{%
  \begingroup
    \everyvbox{}%
    \settoheight\scalemath@axis{$#2\vcenter{}$}%
    \raisebox{\scalemath@axis}{%
      \scalebox{#1}{%
        \raisebox{-\scalemath@axis}{%
          $\m@th#2#3$%
        }%
      }%
    }%
  \endgroup
}
\makeatother 
\newcommand*{\mediumprod}{\scalemath{\mathrel}{.75}{\prod}}
\begin{document}

\title{Time-Slicing Path-integral  in Curved Space}
\author{Mingnan Ding$^{1}$}
\email{dmnphy@sjtu.edu.cn}
\author{Xiangjun Xing$^{1,2,3}$}
\email{xxing@sjtu.edu.cn}
\address{$^1$ Wilczek Quantum Center, School of Physics and Astronomy, Shanghai Jiao Tong University, Shanghai, 200240 China\\
$^2$ T.D. Lee Institute, Shanghai Jiao Tong University, Shanghai, 200240 China \\
$^3$ Shanghai Research Center for Quantum Sciences, Shanghai 201315 China}


\begin{abstract} 
Path integrals constitute powerful representations for both quantum and stochastic dynamics.  Yet despite many decades of intensive studies, there is no consensus on how to formulate them for dynamics in curved space, or how to make  them covariant with respect to nonlinear transform of variables (NTV).  In this work, we construct a rigorous and covariant formulation of time-slicing path integrals for dynamics in curved space.  We first establish a rigorous criterion for equivalence of {\em time-slice Green's function} (TSGF) in the continuum limit (Lemma 1).  This implies the existence of infinitely many equivalent representations for time-slicing path integral.  We then show that, for any dynamics with second order generator, all time-slice actions are equivalent to a Gaussian (Lemma 2).  We further construct a continuous family of equivalent path-integral actions parameterized by an interpolation parameter $\alpha \in [0,1]$ (Lemma 3). We then use these lemmas to develop time-slicing path integral representations both for quantum mechanics and for classical Markov processes in curved space.  The action generically contains a term linear in $\Delta \xv$,  whose concrete form depends on $\alpha$.   Finally we also establish the covariance of our path-integral formalism, by demonstrating how the action transforms under NTV.  The $\alpha = 0$ representation of time-slice action is particularly convenient because it is Gaussian and transforms as a scalar, as long as $\Delta \xv$ transforms according to {\em Ito's formula}. 

\end{abstract}
\maketitle
\section{Introduction}
Path integral as a representation of quantum mechanics was first envisaged by Dirac in his renowned textbook on quantum mechanics~\cite{Dirac(1947)}, and was  developed systematically by Feynman in 1948~\cite{Feynman(1948),Feynman2010}.   Since then, through the hands of many outstanding physicists and mathematicians, it has been transformed into the arguably most powerful tool for theoretical physics~\cite{Kleinert1997}.  The applications of path integral methods range from quantum mechanics~\cite{Justin2010-PI} to quantum field theory~\cite{Zee2010}, quantum open systems~\cite{Feynman2000,Weiss2012}, and quantum gravity~\cite{Hawking-1979-QG}, from Brownian motion to general classical stochastic processes~\cite{Wio2013,Chernyak2006}, as well as polymer physics~\cite{Gennes1979}, and even financial study~\cite{Linetsky1997,Kleinert2009}.  The formalism of path integral not only help shaping our intuition about quantum and classical fluctuations, but also played a key role in the synthesis of quantum field theory with statistical field theory in the last century.   

Consider a one dimensional quantum Hamiltonian $\hat H \!= \! \hat p^2/2 m + V(\hat x)$.  
The Green's function is defined by 
\ba
 i \hbar \frac{\partial}{\partial t} G(x , t |x_0,0)
&=& \hat H G(x , t |x_0,0),  
 \label{tran-amptd-def-1}  \\
G(x , 0 |x_0 ,0) &=& \delta(x  - x_0 ).  
\nonumber
\ea
The path-integral representation of $\!G(x , t |x_0 ,0)\!$  is 
\ba
G(x , t |x_0 ,0)
 =  \int_{(x_0, 0) }^{(x, t)} Dx(t') \, e^{\frac{i}{\hbar} S_{\rm cl} [x(t')]},
\quad \label{PI-quantum-1}
\ea
where the integral means, roughly speaking, summation  over all paths with initial condition $x(0) = x_0 $ and final condition $x(t) = x $, and $S_{\rm cl}[x(t')]$ is the classical action:
\be
S_{\rm cl} [x(t')] = \int_{0}^t \! d t'\, L(x, \dot x)
= \int_{0}^t \! d t' \left[ \frac{m}{2}\dot x^2 - V(x)  \right] .  
\label{classical-action-1}
\ee

\begin{figure}[t]
	\centering
	\includegraphics[width=3in]{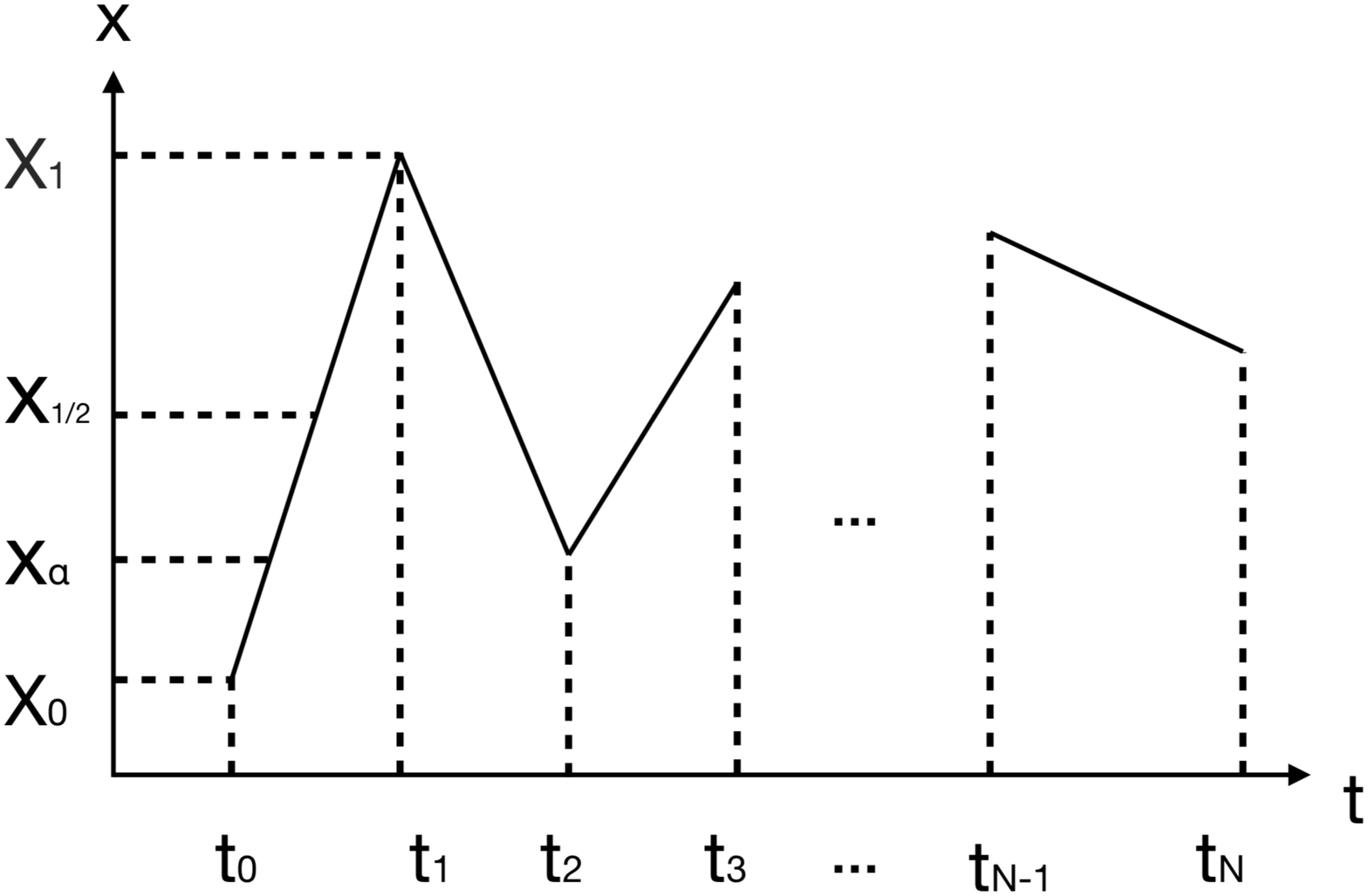}
\vspace{-2mm}
	\caption{Time-slicing: approximation of a continuous path $x(t)$ by a piecewise linear path.  Within each interval $[\xv_k, \xv_{k+1}]$, an intermediate point $\bar \xv_{k}$ needs to be chosen where the potential $V(\xv)$ is evaluated, see Eq.~(\ref{classical-aciton-discretization}).  The choice of $\bar x_k$ is unimportant in flat spaces but important in curved spaces. }
	\label{fig:discretized-path}
\vspace{-3mm}
\end{figure}

To assign a precise meaning to Eq.~(\ref{PI-quantum-1}), one must specify how to sum over paths.  As illustrated in Fig.~\ref{fig:discretized-path}, Feynman  approximated a path $x(t)$ by a sequence of straight-line segments that pass through  $\{x_k = x(t_k ), k = 0, 1, \cdots, N\}$, with $t_k = k \Delta t, \Delta t = t/N,  x_N = x$, and further approximated the  action Eq.~(\ref{classical-action-1}) as a discrete sum:
\be
S_{\rm cl}[x(t')] \approx
\sum_k  { \Delta t} \left[ \frac{m}{2}
\left(\frac{x_{k+1} - x_k}{\Delta t}\right)^2
 - V(\bar x_k) \right],
 \label{classical-aciton-discretization}
\ee
where $\bar x_k$ is a point between $x_k$ and $x_{k+1}$ to be specified.  It will be shown momentarily that in this case the choice of $\bar x_k$ makes no difference in the continuum limit.  The path-integral (\ref{PI-quantum-1}) is then transformed into integration over $N-1$ coordinates $\{x_1, x_2, \cdots, x_{N-1}\}$:
\ba
&& G(x , t |x_0 ,0)  = 
\label{PI-quantum-2} \\
&& C  \!\! \int \! \mediumprod_{k = 1}^{N-1} dx_k \,
 e^{  \frac{i \Delta t}{\hbar}
\sum_k \left[ \frac{m}{2} 
\left(\frac{x_{k+1} - x_k}{\Delta t}\right)^2 
 - V(\bar x_k) \right] } , 
\nonumber
\ea
where the normalization constant $C$ can be fixed by a reference problem, e.g. that of a free particle.  Feynman carried out explicit calculations of Eq.~(\ref{PI-quantum-2}) for free particle and harmonic oscillator, and obtained results consistent with direct solutions to Schr{\"o}dinger equations~\cite{Feynman2010}.  

The prescription for calculation of path integral using Eqs.~(\ref{classical-aciton-discretization}) and (\ref{PI-quantum-2}) is usually called {\em time-slicing}, and is one of many possible {\em regularization schemes} which make the path integral Eq.~(\ref{PI-quantum-1}) finite and calculable.  In field theories, other regularization schemes such as momentum cut-off and dimensional regularization are needed.  We shall not discuss them in this work.     

Because of the imaginary nature of the exponent in Eq.~(\ref{PI-quantum-2}),  the integral is not convergent in the usual sense.   To cure this problem, it is convenient to analytically continuate $t$ into the imaginary axis, $t  \rightarrow - i t$, and transform Eqs.~(\ref{tran-amptd-def-1})  into the {\em imaginary time Schrodinger equation}:
\begin{subequations}
\ba
- \hbar \frac{\partial}{\partial t} G(x , t |x_0,0)
&=& \hat H G(x , t |x_0,0),  
\label{SE-1}\\
G(x , 0 |x_0 ,0) &=&\delta(x  - x_0 ).  
\ea
\label{classical-H-V-1}
\end{subequations}  
 Equation (\ref{PI-quantum-2}) then becomes
\begin{subequations}
 \label{PI-quantum-3}
\ba
G(x , t |x_0 ,0)  =  C  \!\! \int \! \! \mediumprod_{k=1}^{N-1} dx_k \, 
e^ { {-\sum_k A(x_{k+1}, x_k; \Delta t)} }, 
\nonumber\\
\ea  
where $A(x_{k+1}, x_k; \Delta t)$, which shall be called the {\em time-slice action}, is positive and given by
\be
A(x_{k+1}, x_k; \Delta t)=  
{ \frac{\Delta t}{\hbar}} \left[ \frac{m}{2}
\left(\frac{x_{k+1} - x_k}{\Delta t}\right)^2
 \!\! + V(\bar x_k) \right],
 \label{action-A-1}
\ee  
\end{subequations}  
with $\Delta t  = t/N$.  For $V(x) = 0$, Eqs.~(\ref{classical-H-V-1}) reduce to the classical diffusion equation, and Eqs.~(\ref{PI-quantum-3}) define the celebrated {\em Wiener measure}~\cite{wiener1923differential} in the path space of classical Brownian motion.  A mathematically rigorous theory of path integral Eqs.~(\ref{PI-quantum-3}) as a representation of Eqs.~(\ref{classical-H-V-1}) was established by Kac~\cite{Kac1951}. Time-slicing path integral formulation of classical Langevin dynamics was developed by Onsager and Machlup~\cite{OM1953-1,MO1953-2}. 

Let us see why the choice of $\bar \xv_k$  in Eqs.~(\ref{classical-aciton-discretization}) and (\ref{action-A-1}) does not matter.   It is known that for typical paths of classical Brownian motion, $\Delta x \sim \Delta t^{1/2}$, and hence typical paths are continuous but non-differentiable everywhere~\cite{Glimm-Jaffe-book-2012}. It is then evident that the kinetic part and potential part of the total action scale respectively as $\Delta t^{-1}  $ and $\Delta t^0$.  By contrast, the choice of $\bar x_k$ in Eqs.~(\ref{classical-aciton-discretization}) and (\ref{action-A-1}) influences the total action, which is the sum of all time-slice actions, by an amount scaling as $\sum \Delta t \, \Delta \xv \sim \Delta t^{1/2}$, which can be ignored in the continuum limit.

There are however many dynamic processes, either quantum or classical, happening in curved spaces. There are also processes happening in Euclidean space but with curved boundary, and hence are better formulated using curvilinear coordinates.  There are also stochastic processes with multiplicative noises, which resemble very much dynamics in curved spaces.   For these processes, it has been known for long time that the above time-slicing prescription of path integral breaks down.  Consider for example a quantum particle moving in a curved manifold with metric tensor $g_{ij}(\xv)$.  The covariant classical action and covariant quantum Hamiltonian are respectively 
\ba
S_{\rm cl}[\xv(t)] &=& \int dt \left[ \frac{m}{2}g_{ij}
  \dot x^i \dot x^j - V(\xv) \right],
\label{action-curved} \\
\hat H(\xv) &=& - \frac{\hbar^2}{ 2 m \sqrt g}  {\partial_i} 
\sqrt g g^{ij} {\partial_j} 
+ V(\xv),  
\label{Hcurved}
\ea
where repeated indices are summed over, $g^{ij}$ is the inverse matrix of $g_{ij}$, and $g = ( \det g_{ij})= ( \det g^{ij})^{-1}$, all functions of $\xv$. Note that $\partial_i = {\partial}/{ \partial x^i}$ act on everything to the right. It is known that both the classical action and the quantum Hamiltonian transform as scalars under nonlinear transform of variables.
The Green's function (in imaginary time) then is defined by:
\begin{subequations}
\ba
 - \hbar \frac{\partial}{\partial t} G(\xv , t |\xv_0,0)
 &=&\hat H(\xv) \, G(\xv, t | \xv_0, 0), 
 \label{classical-H-V-1-curved-G}
 \\ 
G(\xv, 0 | \xv_0, 0) &=& \ddelta (\xv, \xv_0), 
\ea
\label{classical-H-V-1-curved}
\end{subequations}  
where $\ddelta(\xv, \xv_0)$ is a covariant Dirac delta function, to be defined below in Eq.~(\ref{ddelta-def}).

Following DeWitt's original idea~\cite{Dewitt1957}, many works~\cite{Bastianelli2006,Gervais1976, Salomonson1977, Apfeldorf1996,Karki1997,Boer1996} tried to establish the following path integral representation for Eq.~(\ref{classical-H-V-1-curved})  :
\ba
&& G(\xv, t |\xv_0, 0) \sim C  \!\! \int \! \mediumprod_{k = 1}^{N-1} 
\sqrt{g(\bar \xv_k)} d^d \xv_k \,
\label{interpolation-1}
\\
&& { \exp \! \left\{ \!{- \frac{\Delta t}{\hbar} \! \sum_k \!
\left[  \frac{m}{2} g_{ij}({\bar \xv} _k)
 \frac{\! \Delta x^i_k \Delta x^j_k}{\Delta t^2}
 \!- \! V(\bar \xv_k)  
 \! - \! \delta V(\bar \xv_k ) \! \right]  } \! \right\}  } , \nonumber
\ea
where $\Delta \xv_k \equiv \xv_{k+1} - \xv_k$, and $\delta V(\bar \xv_k ) $ is usually called {\em the  extra term}.  Note, however, unlike the case of Euclidean space, in Eq.~(\ref{interpolation-1}) $\bar \xv_k$ appear both in the kinetic energy and in the integration measure.  Typical variation of $\bar \xv_k$ scales as $\Delta \xv \sim \Delta t^{1/2} $, and leads to change of kinetic energy term scaling as:
\be
\Delta t \sum_k \frac{\Delta \xv^3}{\Delta t^2} 
\sim \sum_k \sqrt{ \Delta t} \sim { \Delta t}^{-1/2},
\ee
which makes a divergent contribution to the action in the continuum limit!  Note that change of $\bar \xv_k$ in the integration measure $\sqrt{g(\bar \xv_k)} d^d \xv_k$ leads to the same effect. Such a correction cannot be compensated by the extra term $\sum_k \delta V(\bar \xv_k ) \Delta t$, which scales as $O(1)$. 
Common  choices for  ${\bar \xv} _k$ are  ${\bar \xv}_k = (1 - \alpha) \xv_k + \alpha \, \xv_{k+1}$, where $\alpha \in[0,1 ]$ shall be called the {\em interpolation parameter} in this work.  Many theories with $\alpha = 0$ (pre-point), $\alpha = 1/2$ (mid-point), and $\alpha = 1$ (post-point) have been formulated, yet the results often do not agree with each other. The in-equivalence between different discretization schemes was pointed out by Langouche {\it et. al.} in \cite{Langouche-book}. 

The above scaling argument already indicates that the action in Eq.~(\ref{interpolation-1}) misses a part linear in $\Delta \xv_k$, summation of which also scales as ${ \Delta t}^{-1/2}$.  This term leads to non-vanishing average of $\Delta \xv_k$, and shall be called  {\em quantum spurious drift}, for reasons that will become clear below.  We will also see that, while the precise form of quantum spurious drift depends on the particular choice of $\alpha$,  for a generic multi-dimensional model, it can not be cancelled by tuning of  $\alpha$, or by a judicial choice of coordinate systems.  

Similar difficulties also arise in path integral representation of classical Markov dynamics in curved spaces, or with multiplicative noises.  Such a dynamics can be described either by the covariant Ito-Langevin equation~\cite{Ding2020,Ding2021}:
\ba
dx^i + \left(  L^{ij} \partial_j U - \frac{1}{\sqrt{g}} \partial_i \sqrt{g} L^{ij} \right) dt
=  b^{i\mu} dW_\mu(t),\nonumber\\
 \label{Langevin-0}
\ea 
with the product in RHS understood in Ito's sense, 
or by the covariant Fokker-Planck equation~\cite{Ding2020,Ding2021}:
\be
\partial_t \, p  = \hat L_{\rm FP} p 
=  \frac{1}{\sqrt{g}} \partial_i \sqrt{g} L^{ij} (\partial_j + (\partial_j U)) p,
\label{FPE-0}
\ee  
where $p(\xv,t) \sqrt{g(\xv)} d^d x$ is the invariant differential probability  of slow variables, and $L_{\rm FP}$ is the invariant Fokker-Planck  operator~\cite{Ding2021}:
\be
\hat L_{\rm FP}  =  \frac{1}{\sqrt{g}} \partial_i \sqrt{g} L^{ij} (\partial_j + (\partial_j U)), 
\label{FPO-def}
\ee 
whereas $ \sum_{\mu}b^{i \mu} b^{j \mu} = L^{ij} + L^{ji}$.     The term $- \frac{1}{\sqrt{g}} \partial_i \sqrt{g} L^{ij}$ in the LHS of Eq.~(\ref{Langevin-0}) is called the {\em spurious drift}, and has been the target of study for long time~\cite{Ding2020,van-Kampen-stochastic,Hanggi-1980,Lau2007,Sasa2017}.  If the space is curved or the coordinates are curvilinear (both imply that $g_{ij}$ depends on $\xv$), or if the noise-amplitudes $b^{i\mu}$ are state-dependent (multiplicative noises), then the same difficulty discussed above also show up, and the corresponding path integral representation becomes much more difficult. 

{Earlier studies of path integral in curved space and in curvilinear coordinates were pursued by DeWitt~\cite{Dewitt1957} and by Edwards and Gulyaev~\cite{Edwards1964}. DeWitt found that the classical action did not lead to correct path integral and tried to introduce an extra term to fix this problem. Edwards and Gulyaev argued that the action of path integral is not invariant under usual rules of calculus when making nonlinear transform of variables (NTV), yet did not supply sufficient details.  Since then, many authors tried to construct path integral representations for quantum and/or classical dynamics in curved space~\cite{Bastianelli2006,Gervais1976,Salomonson1977, Apfeldorf1996,McLaughlin1971,Grosche1987,Hanggi1989,Dekker1976,Fox1986,Durr1978,Alfaro1990,Inoue1985,Grosche1994,Lecheheb2007,Tisza1957,Bhagwat1993,Dekker1979,Janssen1976,Langouche-book}, or to study Langevin dynamics with multiplicative noises~\cite{Tang2014,Arnold2000,Calisto2002,Langouche1978,Langouche-book,Sasa2017,Adib2008,Langouche1978-1}, or to understand how path integral should transform under NTV~\cite{Cugliandolo2017,Cugliandolo2018,Cugliandolo2019}. In spite of the large number of papers published on the subjects, there is still no sign of convergence on opinions.  One school of researchers followed DeWitt's approach and tried to solve the problems by introducing an  extra term, as shown in Eq.~(\ref{interpolation-1})~\cite{Bastianelli2006,Gervais1976,Salomonson1977,Apfeldorf1996}.   Many authors still hold the opinion that actions of path integrals are invariant, or at least can be made invariant via possible revision of time-slicing scheme, under NTV. See, for example, Refs.~\cite{Cugliandolo2017,Cugliandolo2018,Cugliandolo2019} for the most recent effort on one dimensional case.  Comparison of previous works are difficult because of diversity of methods and conventions used, and also because of lack of mathematical details. 

In this work, we shall develop a detailed and systematic construction of time-slicing path integral in curved space, and resolve all the above-mentioned confusing issues. Using asymptotic analysis, we shall first establish three lemmas with reasonable degree of mathematical rigor.  These lemmas constitute the foundations of all later results.  With Lemma 1, we explains why there are so many seemingly different but equivalent representations of time-slicing path integrals, and establish the criterion of equivalence.  With Lemma 2, we prove there is a Gaussian representation of time-slicing path integrals.  With Lemma 3, we explicitly construct a one-parameter family of equivalent representations for the time-slice Green's function (TSGF), parameterized by $\alpha \in [0,1]$.   The time-slicing path integral, whether it is for the quantum or for the classical stochastic process, can then be obtained by straightforward application of these three lemmas.  

For classical Markov processes, we shall further study the connection between time-slicing path integral representation and Langevin dynamics representation.  As is well-known, there are infinite number of equivalent representations for a given Langevin dynamics, each parameterized by a continuous parameter $\bar \alpha$, which specifies where the noise amplitudes are evaluated. The  case of Ito representation ($\bar \alpha = 0$) is particularly simple, because it implies a linear relation between rate of slow variables and noises, conditioned on the slow variables at earlier time.  Exploiting the equivalence between different representations of Langevin dynamics, we shall derive another two-parameter family of equivalent representations for time-slicing path integral of classical Markov processes.  

The last major issue we shall address is the covariance property.  We first clarify how to construct the new time-slice action when a nonlinear transformation of variable is carried out.  We further show that our time-slice action does not transform as a scalar under usual transformation rules of tensor algebra, mainly due to the non-differentiable nature of typical paths that dominate path integrals.  Finally, we explicitly show that the $\alpha = 0$ time-slice action can be made invariant if we demand that $\Delta \xv$ transform according to Ito's formula. 

The remaining of this work is organized as follows.  In Section \RNum{2},  we establish three lemmas, which constitute  the base of all  later results.  In Section \RNum{3}, we apply these lemmas to obtain a continuous family of equivalent representations for time-slicing path integral.   We also revisit the problem of Edwards and Gulyaev and discuss the geometric origin of spurious drift.  In Section \RNum{4}, we study the connection between classical nonlinear Langevin dynamics and time-slicing path integral, and obtain a more general family of equivalent path integral representations for classical Markov processes.  In Section \RNum{5}, we discuss the covariance property of time-slicing path integral.  Finally, in Section \RNum{6}, we draw the concluding remark and outline future research directions.  In Appendix A, we supply a detailed proof of Lemma 3.

\section{ {Three Lemmas about TSGF} }
Consider a Riemannian manifold with coordinate system $\xv = \{x^1,x^2,\cdots,x^d \}$ and metric tensor $g_{ij}(\xv)$, and $g(\xv) = \det (g_{ij}(\xv))$. We define the volume measure $d\mu(\xv) $ and the invariant volume measure $dv(\xv)$ as
\begin{subequations}
\label{dmu-dv-def}
\ba
d\mu(\xv) &\equiv& dx^1 dx^2 \cdots dx^d,
\label{d-mu-def} \\
dv(\xv) &\equiv& \sqrt{g(\xv)} \, d\mu(\xv). 
\label{d-v-def}
\ea 
\end{subequations}
We also define the covariant Dirac delta function as
\ba
\ddelta ( \xv, \xv_0) \equiv \frac{1}{\sqrt {g(\xv_0)}}
\prod_{i=1}^d   \delta(x^i -  x_0^i)  
= \ddelta ( \xv_0, \xv),
\label{ddelta-def}
\ea
where $ \delta(x^i -  x_0^i)$ is the usual 1d Dirac delta function.  For an arbitrary function $f(\xv)$ on the manifold, we have
\be
\int f(\xv ) \ddelta  ( \xv, \xv_0)  dv(\xv) = f(\xv_0). 
\ee

We consider a dynamics generated by a second order partial differential operator:
\ba
\hat L(\xv) &\equiv& 
\frac{1}{\sqrt{g(\xv)} } \partial_i\partial_j \sqrt{g(\xv)} D^{ij} (\xv)
 \nonumber\\ 
&-& \frac{1}{\sqrt{g(\xv)} } \partial_i \sqrt{g(\xv)} F^i(\xv)  - \Phi (\xv), 
\label{DL}
\ea
where $D^{ij}(\xv) = D^{ji}(\xv) $ is symmetric. For now we assume that $D^{ij}(\xv)$  is non-singular, and that $D^{ij} (\xv),  F^i(\xv), \Phi (\xv) $ do not depend on time.  Generalization to the case where these functions are time-dependent is straightforward. The case of singular $D^{ij}(\xv)$ is more complicated, and shall not be discussed in this work. It is clear that $\hat L(\xv) $  as defined in Eq.~(\ref{DL}) contains as special cases both the quantum Hamiltonian Eq.~(\ref{Hcurved}) and the classical Fokker-Planck operator Eq.~(\ref{FPE-0}).  


The Green's function for Eq.~(\ref{DL}) is then defined as 
\be
G(\xv , t| \xv_0, 0 )  = e^{\hat L(\xv) t} \ddelta(\xv, \xv_0 ),
\label{P-transition-0}
\ee 
which also contains the Green's function defined in Eqs.~(\ref{classical-H-V-1}) and (\ref{classical-H-V-1-curved}) as special cases.   $G(\xv , t| \xv_0 , 0 )$ is also known as {\em propagator, transition amplitude} (quantum mechanics), or {\em transition probability} (classical stochastic processes) etc.   We shall first develop a path integral representation of Eq.~(\ref{P-transition-0}), and then apply it to the specific cases of quantum mechanics and classical stochastic dynamics.


As illustrated in Fig.~\ref{fig:discretized-path}, we cut the time interval $[0,t]$ into $N$ slices with duration $\Delta t = t/N$, and introduce TSGF:
\be
G(\xv |   \xv_0 ; \Delta t ) = e^{\hat L(\xv) \Delta t} \ddelta(\xv, \xv_0),
\label{P-transition-0-infini}
\ee 
with $\Delta t = t/N$.  Equation (\ref{P-transition-0}) can be rewritten into a product of $N-1$ consecutive TSFGs:
\ba
G(\xv , t| \xv_0 , 0 )
&=& \int dv_{N -1} \cdots dv_1 \, 
 G(\xv_{N} |\xv_{N-1}; \Delta t) 
 \nonumber\\
&& \,\, \cdots  G(\xv_{2} | \xv_{1} ; \Delta t )
 G(\xv_{1} | \xv_{0} ; \Delta t) , \quad
 \label{P-PPPPP-1}
\ea
where $\xv_{N} = \xv$, and $dv_k = dv(\xv_k)$.  The discretized path $(\xv_N, \xv_{N -1}, \ldots, \xv_1, \xv_0)$ forms {\em the cylinder set}~\cite{Glimm-Jaffe-book-2012}, which can be used to construct rigorously the sigma algebra of the path space.  

To construct path integral representation of Green's function, we approximate each TSGF in the RHS  of Eq.~(\ref{P-PPPPP-1}) by an exponential form.  The approximation should be such that in the {\em continuum limit} $\Delta t \rightarrow 0$, the exact Green's function is recovered.  More concretely, all moments of the approximated Green's function, i.e., the averages of powers of $\Delta \xv =  \xv - \xv_0$ calculated using the approximated version of RHS of Eq.~(\ref{P-PPPPP-1}), must give the exact answers in the limit $\Delta t \rightarrow 0$.  Any such approximation of TSGF shall be called {\em a representation}.   Now recall that the number  $N$ of steps is related to the duration $\Delta t$ of each step via $\Delta t = t/N$.  If the error in each step is much smaller than $\Delta t$, then the total error in the entire interval is much smaller than $N \times \Delta t = t$, and hence becomes negligible in the continuum limit.  Likewise, if two approximations of $G( \xv_{k+1} | \xv_k ; \Delta t )$ yield the same moments up to order of $\Delta t$, they become indistinguishable in continuum limit, and we shall call these two approximations {\em equivalent}.  It then follows that all representations of TSGF are equivalent to each other.   Hence we obtain the first lemma: 
\begin{lemma} \label{lemma-1} 
An approximation of TSGF is a representation if and only if it yields correct moments up to $\Delta t$.  Two approximations of TSGF are equivalent if they yield the same moments  up to $\Delta t$.  
\end{lemma}

As we will demonstrate, there are infinitely many seemingly different but equivalent representations of TSGF, each of them can be used to construct the time-slicing path integral representation of Green's function.  Lemma 1 gives a concrete criterion for testing of the correctness of various approximations, and of the equivalence between different representations  of TSGF.  Whilst the content of Lemma 1 was recognized or used implicitly by many authors previously, it is important to make it explicit, as it will play an essential role in the analyses below.

Let $\Delta \xv =  \xv - \xv_0$, and $\Delta x^i$ the i-th component of $\Delta \xv$.  A $M$-th moment of TSGF, Eq.~(\ref{P-transition-0-infini}), is defined as 
\begin{subequations}
\ba
&& \langle \Delta x^{i_1} \cdots \Delta x^{i_M}  \rangle 
\nonumber\\
&=& \!\! \int \!\! dv(\xv) \Delta x^{i_1} \cdots
 \Delta x^{i_M} G (\xv | \xv_0; \Delta t ),
\label{moments-def}
\ea 
 The operator $e^{\hat L(\xv) \Delta t}$ in Eq.~(\ref{P-transition-0-infini}) may be expanded in terms of $\Delta t$.  According to Lemma 1, the expansion can be truncated at the first order: 
\ba 
&& \langle \Delta x^{i_1} \cdots \Delta x^{i_M}  \rangle 
\label{moments-def-1}
\\
&=& \!\!  \int \!\! dv(\xv)
\Delta x^{i_1} \cdots \Delta x^{i_M}  
\! \left[ 1 + \Delta t \, \hat L(\xv) 
 + \cdots \right] \!  \ddelta (\xv, {\mathbf \xv_0}). 
\nonumber
\ea
Using Eqs.~(\ref{DL}), (\ref{dmu-dv-def}), and (\ref{ddelta-def}), and integrating by parts, we see that the lowest three moments are respectively: 
\ba
\langle 1 \rangle &=&  1 - \Phi(\xv_0) \Delta t + O(\Delta t^2), 
\nonumber\\
\langle \Delta x^i \rangle &=& F^i(\xv_0) \Delta t + O(\Delta t^2),
\label{three-moments-0}\\
\langle \Delta x^i \Delta x^j  \rangle &=&
2  D^{ij}(\xv_0) \Delta t + O(\Delta t^2).
\nonumber
\ea
\end{subequations}
For all $M\geq 3$, the integral in Eq.~(\ref{moments-def-1}) vanishes identically (up to $\Delta t$) because of the Dirac delta function. Hence  we arrive at the second lemma:

\begin{lemma} \label{lemma-2} 
For all $M\geq 3$, $M$-th order moments  of TSGF Eq.~(\ref{P-transition-0}) are at least of order $\Delta t^2$, and hence makes no contribution in the continuum limit. 
\end{lemma} 
\vspace{-2mm}

Now consider the following Gaussian distribution: 
\ba
\frac{ d\mu(\xv)\, e^{- (\Delta x^i - F^i(\xv_0) \Delta t)
\frac{D^{-1}_{ij}(\xv_0)}{4 \Delta t}
 (\Delta x^i - F^i(\xv_0) \Delta t) - \Phi(\xv_0) \Delta t} }
{\sqrt{(4\pi \Delta t)^d \det D^{ij}(\xv_0) }} ,\nonumber
 \nonumber\\
 \label{Wissel-alpha-0-main-1}
\vspace{-3mm}
\ea
where $D^{-1}_{ij}$ is the inverse matrix of $D^{ij}$, and $\det D^{ij}$ is the determinant of the matrix $D^{ij}$.  Note that all functions are evaluated at $\xv_0$, and we are attaching the volume element in Eq.~(\ref{Wissel-alpha-0-main-1}).  It is clear that the three lowest order moments of Eq.~(\ref{Wissel-alpha-0-main-1}) are those given in Eq.~(\ref{three-moments-0}), and all higher order moments are at least of order $\Delta t^2$.  Then according to Lemma 1, Eq.~(\ref{Wissel-alpha-0-main-1}) is a representation of TSGF, and hence can be used to construct the time-slicing path integral.   We shall call Eq.~(\ref{Wissel-alpha-0-main-1}) the {\em Gaussian representation} of TSGF. 

Whilst lemmas 1 and 2 are easy to establish, their importance can hardly be overrated.  They constitute a starting point for a systematic construction of equivalent representations of time-slicing path integrals.   In particular, to verify that certain approximation of TSGF is a representation, i.e., it can be used to construct time-slicing path integral, we only need to show that (i) its lowest three moments are the same as those of Eq.~(\ref{Wissel-alpha-0-main-1}), and (ii) all higher moments are smaller than $\Delta t$.  
For a discussion of Gaussian representation of TSGF from the viewpoint of renormalization group, see Ref.~\cite{Janssen1992}.  

Using these results, we can construct a continuous family of equivalent representations of TSGF that is parameterized by an interpolation parameter $\alpha \in [0,1]$.  This is the content of Lemma 3:

\vspace{-2mm}
\begin{widetext}
\begin{lemma}\label{lemma-3}
Let $D^{ij}(\xv) = D^{ji}(\xv) $ be symmetric and positive, $F^i(\xv)$ a vector,  and $\Phi(\xv)$ a scalar, all functions of $\xv$.  Let $\Delta \xv = \xv - \xv_0$, and $\xv_{\alpha} = \xv_0 + \alpha \Delta \xv$ with $\alpha \in [0,1]$. The following one-parameter family of distributions are equivalent to each other, in the sense that   their moments are all equal up to order $\Delta t$:
\vspace{-2mm}
\ba
&& \!\!\!\! \frac{d \mu(\xv) }
{ \sqrt{ (4\pi \Delta t)^d \det D^{ij}(\xv_{\alpha} )}}
 \exp \bigg\{ \!
- \! \Big( \Delta x^i -  F^i(\xv_{\alpha})  \Delta t
 + 2 \alpha \partial_k D^{ik}(\xv_{\alpha})  \Delta t \Big)
 \frac{D_{ij}^{-1}(\xv_{\alpha}) }{4\Delta t}
  \Big( \Delta x^j -  F^j(\xv_{\alpha})  \Delta t
  + 2 \alpha \partial_l D^{jl}(\xv_{\alpha}) \Delta t \Big)
\nonumber \\ 
&& \hspace{43mm}  - \,  \alpha \, \partial_i F^i  (\xv_{\alpha} )  \Delta t  
 + \alpha^2 \partial_i \partial_j D ^{ij}  (\xv_{\alpha} ) \Delta t
 - \Phi (\xv_{\alpha} ) \, \Delta t  \bigg\}. 
\label{Wissel-main-0}
\ea
\end{lemma}
\vspace{-3mm}
Note that all functions are evaluated at $\xv_{\alpha}$.

There is an extension of Lemma 3, which provides more flexibility in implementation of time-slicing.  Using similar methods, it can be proved that the following two-parameter family of equivalent representations, where $D^{ij}$ is evaluated at $\xv_{\alpha_1} = \xv + \alpha_1 \Delta \xv $, and whereas $F^i$ is evaluated at $\xv_{\alpha_2} = \xv +\alpha_2  \Delta \xv$, whilst $\Phi $ can be evaluated at arbitrary place:
\ba
\vspace{3mm}
&& \frac{d \mu(\xv) }{\sqrt{(4 \pi \Delta t)^{d} \det D^{ij} ( \xv_{\alpha_1})} } 
 \exp \bigg\{ 
 \nonumber\\
 &-  &  \bigg( \Delta x^i  - F^i (\xv_{\alpha_2} )   \Delta t 
+ 2 \alpha_1 \partial_k D^{ik} (\xv_{\alpha_1})   \Delta t  \bigg)
\frac{D^{-1}_{ij}(\xv_{\alpha_1}) }{4 \Delta t} 
 \bigg( \Delta x^j  - F^j (\xv_{\alpha_2})  \Delta t  
 + 2 \alpha_1 \partial_l D^{jl} (\xv_{\alpha_1} )
   \Delta t \bigg)
\nonumber \\
 &-&  \alpha_2 \partial_i F^i  (\xv_{\alpha_2} )  \Delta t  
 + \alpha_1^2 \partial_i \partial_j D ^{ij}  (\xv_{\alpha_1} ) \Delta t
 - \Phi \Delta t   \bigg\}.  
 \label{trans-amp-classical-neq-2}
\ea
 
 The proof of Lemma 3 is technically very complicated, and is presented in Appendix \ref{transitionformula}.  Historically, this equivalent class was obtained by Wissel in 1979~\cite{Wissel1979}.  Wissel's work however has not received much attention, most likely due to its lack of mathematical rigor.  Our proof of Eq.~(\ref{Wissel-main-0}) is systematic and rigorous.  For 1d case, Eq.~(\ref{Wissel-main-0}) reduces to the result of Haken~\cite{Haken1976}.  
 
Several important comments are in order.  Firstly note that the $\alpha = 0$ version of Eq.~(\ref{Wissel-main-0}) is precisely the Gaussian representation Eq.~(\ref{Wissel-alpha-0-main-1}).  Secondly for $\alpha \neq 0$, Eq.~(\ref{Wissel-main-0}) is {\em not} Gaussian in $\Delta \xv$, due to the hidden dependence of various functions on $\Delta \xv$.  Thirdly, there are (infinitely many) other representations of TSGF that do not assume the form of Eq.~(\ref{Wissel-main-0}).  For example, the factor $\det D^{ij}$ in front of the exponential may be evaluated at a  point different from $\xv_{\alpha}$.  This leads to further revision of the time-slice action.  We shall not explore this issue further.

\section{The $\alpha$-Representation}
\label{sec:alpha-rep}
\vspace{-3mm}

We will now study  time-slicing path integral representation in curved space, first for quantum mechanics, and then for classical Markov processes.  We shall follow the procedure outlined in the preceding section, first derive a Gaussian representation, then construct arbitrary $\alpha$-representation with $\alpha \in [0,1]$.  These representations are equivalent to each other, as guaranteed by Lemma 3.  

\vspace{-3mm}
\subsection{Quantum mechanics in curved space}

The imaginary time Green's function is already defined in Eqs.~(\ref{classical-H-V-1-curved}).   The TSGF can be written as 
\be
G(\xv | \xv_0; \Delta t )= 
e^{-\Delta t \hat H(\xv)/\hbar } \ddelta(\xv, \xv_0), 
\label{G-quantum-def}
\ee
where $\hat H$ is given in Eq.~(\ref{Hcurved}).   The moments of $G(\xv | \xv_0; \Delta t ) $ are calculated up to the order of $\Delta t$, by expanding the exponential and using integration by parts:
\begin{subequations}
\label{ave-moments-dx}
\ba
\langle 1 \rangle 
&=&  \! \! \int  d \mu(\xv)  \, \sqrt{g(\xv)} \,\,
 e^{-\Delta t \hat H(\xv)/\hbar } \ddelta(\xv, \xv_0)
=  1 -  \frac{\Delta t}{\hbar} V(\xv_0) + O(\Delta t^2),
\\
\langle \Delta x^k \rangle &=&  \!\!
\int d \mu(\xv)  \, \sqrt{g(\xv)} \Delta x^k
 e^{-\Delta t \hat H(\xv)/\hbar } \ddelta(\xv, \xv_0)
=  \frac{ \hbar \Delta t}{2 m} \left( \frac{1}{\sqrt{g}} 
\partial_j \left(\sqrt{g }g^{jk} \right) \right)_{\!\! 0}
+ O(\Delta t^2),
 \label{ave-dx-quantum}  \\
\langle \Delta x^k \Delta x^l  \rangle &=&  \!\!
\int d \mu(\xv)  \, \sqrt{g(\xv)} \Delta x^k \Delta x^l
 e^{-\Delta t \hat H(\xv)/\hbar } \ddelta(\xv, \xv_0)
=  \frac{\hbar \Delta t}{m}\, g^{ij}(\xv_0) + O(\Delta t^2),
\ea
\end{subequations}
where $(\cdots)_{0}$ means that all functions inside the bracket are evaluated at $\xv_{0}$. 

Using these moments, we can construct a Gaussian expression for the TSGF:
\begin{subequations}
\label{trans-amp-quantum}
\ba
dv(\xv )\, G(\xv | \xv_0; \Delta t ) = 
\frac{ \sqrt{g(\xv_0)}\, d \mu(\xv) }{(2 \pi \hbar \Delta t/m)^{d/2}}
e^{- A^0 (\xv, \xv_0; \Delta t)},  
\label{trans-amp-quantum-1}
\ea
where the {\em time-slice action} $A^0(\xv, \xv_0; \Delta t) $ is given by
\ba
  A^0 (\xv, \xv_0; \Delta t) &=& 
   \left[  \Delta x^i  -  \frac{\hbar \Delta t}{2 m } 
\left(  \frac{1}{\sqrt{g}}\, \partial_k  \sqrt{g}  g^{ki} \right)_{\!\! 0} \, \right] 
\frac{g_{ij}(\xv_0)}{2 \hbar \Delta t /m}
\left[   \Delta x^j -   \frac{\hbar \Delta t}{2 m} 
\left( \frac{1}{\sqrt{g}}\,\partial_l \sqrt{g} \,  g^{jl} \right)_{\!\! 0}  \, \right] 
+ \frac{\Delta t}{\hbar} V(\xv_0),
\label{trans-amp-quantum-2}\\
&=& \left[   \Delta x^i 
+  \frac{\hbar \Delta t}{2m }  g^{kl}(\xv_0) \Gamma^{i}_{kl} (\xv_0)  \right] 
\frac{g_{ij}(\xv_0) }{2 \hbar \Delta t /m}
\left[  \Delta x^j + \frac{\hbar \Delta t}{2 m} 
  g^{mn}(\xv_0) \Gamma^{j}_{mn}(\xv_0)  \right] 
+ \frac{\Delta t}{\hbar} V(\xv_0),
\label{trans-amp-quantum-2-1}
\ea
\end{subequations}
where all functions are evaluated at $\xv_0$, and $\Gamma^{j}_{mn}$ is the Christoffel symbol, constructed from the metric tensor:
\begin{subequations}
\ba
\Gamma^k_{ij} &\equiv& \frac{1}{2} \, g^{kl} ( \partial_i g_{jl}
 + \partial_j g_{il} - \partial_l g_{ij}),
\vspace{-2mm}
\ea  
and we have used the contracting relations: 
\ba
 g^{kl} \Gamma^i_{kl} &=& - \frac{1}{\sqrt{g}} \partial_k (\sqrt{g} g^{ik}). 
\ea
\vspace{-2mm}
\end{subequations}

The distribution Eq.~(\ref{trans-amp-quantum-1}) is Gaussian, because the action $A^0 (\xv, \xv_0; \Delta t)$ is quadratic in $\xv$, and the prefactor $ \sqrt{g(\xv_0)}\, d \mu(\xv)$ is independent of $\xv$.  One interesting feature about Eq.~(\ref{trans-amp-quantum-2}) is that  the average of $\Delta \xv$ is non-vanishing if the metric tensor is not constant.  This is what we call the {\em quantum spurious drift}, which has been missed by many previous studies on path integral representation of quantum mechanics in curved space.

Now let us compare Eqs.~(\ref{trans-amp-quantum-2}) with (\ref{Wissel-alpha-0-main-1}), and make the identification $F^i = ({\hbar}/{(2 m \sqrt{g}}))  \partial_k (\sqrt{g }g^{ki} ) $, $D^{ij} = \hbar \,g^{ij}/(2 m)$, and $\Phi = V/\hbar$.  Applying Lemma 3, we find the following one-parameter family of representations for TSGF, all of which equivalent to Eqs.~(\ref{trans-amp-quantum}):
\begin{subequations}
  \label{qmalpha-all}
\ba
dv(\xv)\, G(\xv | \xv_0; \Delta t ) &=& 
\frac{\sqrt{g(\xv_{\alpha})}\, d \mu(\xv) }{(2 \pi \hbar \Delta t /m)^{d/2}}
e^{- A^\alpha (\xv, \xv_0; \Delta t)},  
  \label{qmalpha-0-0}
\ea
\ba
A^\alpha (\xv, \xv_0; \Delta t) &=&  
\left[  \Delta x^i  
 - \frac{\hbar \Delta t}{m}  \left( \frac{1}{2\sqrt{g}}\partial_k \sqrt{g }g^{ik}
 -     \alpha \partial_k g^{ik} \right)_{\!\!\alpha} \right]
\frac{g_{ij}(\xv_{\alpha})}{2 \hbar \Delta t /m}
\left[ \Delta x^j  -  \frac{\hbar \Delta t}{m}  \left( \frac{1}{2\sqrt{g}}\partial_l \sqrt{g }g^{jl}
 -     \alpha \partial_l g^{il} \right)_{\!\!\alpha}  \right] 
\nonumber \\ 
& +& \frac{\hbar \alpha}{2 m} \Delta t \bigg(\partial_i  \sqrt{{ 1}/{g}} \, 
\partial_k \sqrt{g }g^{ki}  \bigg)_{\!\!\alpha}
-  \frac{\hbar \alpha^2}{2 m} \Delta t (\partial_i \partial_j g^{ij})_{\alpha}
  +  \frac{\Delta t}{\hbar} V(\xv_{\alpha}).
  \label{qmalpha-0}
\ea
\end{subequations}
where $(\cdots)_{\alpha}$ means that all functions are evaluated at $\xv_{\alpha}$.  
It is seen from Eq.~(\ref{qmalpha-0}) that whilst the detailed form of quantum spurious drift depends on the choice of $\alpha$, it is always non-vanishing for a generic multi-dimensional model.  The lesson we learn here is that a typical quantum trajectory in curved space or in curvilinear coordinates behaves as a biased random walk.  For the special case $\alpha = 1/2$, the quantum spurious drift reduces to 
\ba
&&\frac{ \hbar}{m} \left( \frac{1}{2\sqrt{g}}\partial_k \sqrt{g }g^{ik} 
- \alpha \partial_k g^{ik}  \right)
\rightarrow \frac{\hbar}{2 m } g^{ik} \partial_k \log \sqrt{g} 
= \frac{\hbar}{2 m} g^{ik} \Gamma^j_{kj}.  
\ea

\subsection{Classical Markov processes}
Let us now apply the same procedure to the classical case, whose TSGF is
\be
G(\xv | \xv_0; \Delta t ) = e^{-\Delta t \hat L_{\rm FP}(\xv) } \ddelta(\xv, \xv_0), 
\ee
where the Fokker-Planck operator $ \hat L_{\rm FP}(\xv)$ is given in Eq.~(\ref{FPO-def}).  The first three moments can be calculated straightforwardly. Ignoring higher order terms, we have
\begin{subequations}
\ba
\langle 1 \rangle \!\! &=&  1 + O(\Delta t^2), 
\label{moment-0-classical} \\
 \langle \Delta x^k \rangle
&=& \Delta t\! \left(  \frac{1}{\sqrt{ g }} \partial_j 
\left( \sqrt{g} L^{kj} \right) 
- L^{kj}\partial_j U \right)_{\!\! 0}
+ O(\Delta t^2), \\
\langle \Delta x^k \Delta x^l  \rangle 
&=&  2 \Delta t\, B^{kl}(\xv_0) + O(\Delta t^2),
\ea
\end{subequations}
where $(\cdots)_{0}$ means that all functions inside the bracket are evaluated at $\xv_{0}$. The matrix $B$ is the symmetric part of $L$, i.e., $2 B^{ij} = L^{ij}+ L^{ji}$. For classical Markov processes, $G(\xv | \xv_0; \Delta t )$ is interpreted as the transition probability.  Hence Eq.~(\ref{moment-0-classical}) can be understood as normalization of total probability.

Using these moments, we obtain a Gaussian representation for  TSGF:
\ba
&& dv(\xv) \, G(\xv | \xv_0; \Delta t ) = 
\frac{d \mu(\xv)  \, e^{-   A^0 (\xv, \xv_0; \Delta t)}}
{\sqrt{(4 \pi \Delta t)^{d} \det B^{ij}(\xv_0)} } ,  \quad 
\label{trans-amp-classical-1}
\ea
where the time-slice action $A^0(\xv , \xv_0; \Delta t) $ is:
\ba
&& A^0(\xv , \xv_0; \Delta t)  =
 \bigg[ \Delta x^i - \Delta t \left( \frac{1}{\sqrt{ g }} \partial_k 
\left( \sqrt{g} L^{ik} \right) 
- L^{ik}\partial_k U \right)_{\!\! 0}   \bigg] 
\frac{ B^{-1}_{ij}(\xv_0)}{4 \Delta t} 
 \bigg[  \Delta x^j  - \Delta t \left( \frac{1}{\sqrt{ g }} \partial_l 
\left( \sqrt{g} L^{jl} \right) 
- L^{jl}\partial_l U \right)_{\!\! 0}  \bigg] ,
\label{trans-amp-classical-2}
\label{trans-amp-classical}  \quad\quad
\quad
\ea
where $(B^{-1})_{ij}$ is the inverse matrix of $B^{ij}$, and  all functions are evaluated at the initial point $\xv_0$.  

Invoking Lemma 3, we obtain a one-parameter family of equivalent representations: 
\begin{subequations}
\label{trans-amp-classical-neq-0}
\ba
dv(\xv) G(\xv | \xv_0; \Delta t ) &=& 
\frac{d \mu(\xv) \, e^{- A^\alpha (\xv, \xv_0; \Delta t)}}
{\sqrt{(4 \pi \Delta t)^{d} \det B^{ij}(\xv_{\alpha} )} } ,
\label{trans-amp-classical-neq-0-0} \\
A^\alpha(\xv , \xv_0; \Delta t) &=&  \bigg[  \Delta x^i 
-  \Delta t \left( F^i   - 2 \alpha \partial_k B^{ik}  \right)_{\! \alpha} \bigg]
\frac{B^{-1}_{ij}(\xv_{\alpha}) }{4 \Delta t} 
 \bigg[ \Delta x^j  -  \Delta t  \left( F^j 
  - 2 \alpha \partial_l B^{jl}  \right)_{\alpha} \bigg] 
 \nonumber \\
 &+ & \alpha \,  ( \partial_i F^i )_{ \alpha}  \Delta t 
  - \alpha^2 ( \partial_i \partial_j B ^{ij} )_{\alpha} \Delta t,  
\label{trans-amp-classical-action-neq-0}
\ea
\end{subequations}
where all functions are evaluated at $\xv_{\alpha} = \xv + \alpha \Delta \xv$, and $F^i$ is 
\ba
&& F^i =  \Delta t {\sqrt{1/g}} \,\left( \partial_k \sqrt{g}L^{ik} \right)
 - \Delta t\, L^{ik} \partial_k U. 
\ea 

More generally, we can apply the extension of Lemma 3, and obtain a two-parameter family of equivalent representations of TSGF:
\ba
&&  dv(\xv) \, G(\xv | \xv_0; \Delta t ) = 
\frac{d \mu(\xv)  \, e^{- A^{\alpha_1, \alpha_2}(\xv, \xv_0; \Delta t)}}
{\sqrt{(4 \pi \Delta t)^{d} \det B^{ij}( \xv_{\alpha_1} ) } } , 
\nonumber\\
&& A^{\alpha_1, \alpha_2} (\xv , \xv_0; \Delta t) =  
  \alpha_2 \partial_i F^i  (\xv_{\alpha_2} )  \Delta t  
 - \alpha_1^2 \partial_i \partial_j B ^{ij}  (\xv_{\alpha_1} ) \Delta t
\label{trans-amp-classical-neq-0}
 \\ 
 &+& \bigg( \Delta x^i  - F^i (\xv_{\alpha_2} )   \Delta t 
+ 2 \alpha_1 \partial_l B^{il} (\xv_{\alpha_1})   \Delta t  \bigg)
\frac{B^{-1}_{ij}(\xv_{\alpha_1}) }{4 \Delta t} 
 \bigg( \Delta x^j  - F^j (\xv_{\alpha_2})  \Delta t  
 + 2 \alpha_1 \partial_m B^{jm} (\xv_{\alpha_1} )
   \Delta t \bigg),
\nonumber 
\ea
where $B^{ij}$ is evaluated at $\xv_{\alpha_1} = \xv + \alpha_1 \Delta \xv $, and $F^i$ at $\xv_{\alpha_2} = \xv +\alpha_2  \Delta \xv$.

\vspace{3mm}

\end{widetext}

\subsection{Two Examples}

\subsubsection{The problem of Edwards-Gulyaev}
\label{sec:EG-1}
We consider the problem studied by Edwards and Gulyaev~\cite{Edwards1964}: a free particle moves in a flat plane using polar coordinates.  The quantum Hamiltonian is
\ba
\hat H &=& -\frac{1}{2} (\partial_x^2 + \partial_y^2)
\nonumber\\
&=& - \frac{1}{2} \left( 
\frac{1}{r} \partial_r r \partial_r + \frac{1}{r^2} \partial_\phi^2
\right),
\label{H-2d-polar}
\ea
where we have set $\hbar = m = 1$.  The classical action is
\begin{subequations}
\ba
S_{\rm cl} &=& \frac{1}{2} \int dt  (\dot x^2 + \dot y^2) 
\label{S-2d-Cartesian}\\
&=& \frac{1}{2} \int dt ( \dot r^2 + r^2 \dot \phi^2),
\label{S-2d-polar-1}
\ea
\end{subequations}
which transforms as a scalar under NTV.   In Cartesian coordinates, the parameter $\alpha$ is irrelevant. The TSGF is 
\ba
dxdy \, G(x,y | x_0,y_0; \Delta t ) = \frac{ dxdy \, e^{ 
- ({ \Delta x^2 + \Delta y^2})/(2 \Delta t)}}{2\pi  \Delta t},\nonumber\\
\label{tpcc}
\ea
where $\Delta x = x - x_0 , \Delta y = y - y_0 $.  The time-slice action then reads
\be
A(\Delta x, \Delta y; \Delta t) = 
\frac{ \Delta x^2 + \Delta y^2} { 2 \Delta t}. 
\label{action-EG-Cartesian}
\ee


 Using polar coordinates:
 \ba
 r = \sqrt{x^2+y^2}, &&  \phi = \arctan(y/x),
 \label{polar-transform}\\
  r_0 = \sqrt{x_0^2+y_0^2},  && \phi_0 = \arctan(y_0/x_0),
 \ea 
Edwards and Gulyaev transform Eq.~(\ref{tpcc}) into
\ba
&&  dv(r, \phi) \,  G(r, \phi| r_0, \phi_0; \Delta t)
\nonumber\\
&=&  \frac{ dv(r, \phi)  \, e^ { - [ { r^2 +r_0^2
- 2 r r_0 \cos ( \phi - \phi_0 )}]/ (2 \Delta t) }  }{2\pi \Delta t},
\label{discrete-polar}
\ea
where $dv(r, \phi)  = r dr d \phi = dx dy$.  Defining $\Delta r = r - r_0, \Delta \phi = \phi - \phi_0$, and realizing that $\Delta r, \Delta \phi \sim \sqrt{\Delta t}$, the negative exponent of Eq.~(\ref{discrete-polar}) can be expanded up to order of $\Delta t$:
\ba
&&  \frac{  r^2 +r_0^2
- 2 r r_0 \cos ( \phi - \phi_0 ) }{ 2 \Delta t} =
\label{discrete-polar-2}
\\
&&\frac{1}{2 \Delta t} \left[ 
\Delta r^2 + r_0^2 \Delta \phi^2 
+ r_0 \Delta r \Delta \phi^2 
- \frac{r_0^2 \Delta \phi^4}{12} 
\right] + O(\Delta t^{\frac 3 2}).
 \nonumber 
\ea
If we only keep the first two terms in the square bracket in RHS of Eq.~(\ref{discrete-polar-2}), we would obtain $ \left( \Delta r^2 + r_0^2 \Delta \phi^2 \right)/{2 \Delta t}$, which is the counterpart of the classical action (\ref{S-2d-polar-1}) of a small step $\Delta t$. However, because $\Delta r, \Delta \phi \sim \sqrt{\Delta t}$, the third and forth terms in the square bracket scale respectively as $\Delta t^{1/2}$ and $\Delta t$ (taking into account the factor $1/\Delta t$ outside the bracket), and hence can not be ignored according to Lemma 1.  Edwards and Gulyaev noticed~\cite{Edwards1964} the importance of the fourth term, but missed the third term.

This problem can be easily solved using our method.  We can treat it as either a quantum case or a classical case.  As a quantum case, we note that the polar coordinate version of  Eq.~(\ref{H-2d-polar}) can be written as Eq.~(\ref{Hcurved}) with $\hbar = m =  1$ and $V = 0$, and 
\be
g_{ij} = \begin{pmatrix} 
1 & 0\\ 0 & r^2 \end{pmatrix}, \quad
g^{ij} = \begin{pmatrix} 
1 & 0\\ 0 & r^{-2} \end{pmatrix},
\quad g(r) = r^2.  
\label{metric-2d-polar}
\ee
Substituting these back into Eqs.~(\ref{trans-amp-quantum-1}) and (\ref{trans-amp-quantum-2}), we find the TSGF:
\ba
dv(r, \phi)  \, G(r, \phi | r_0, \phi_0; \Delta t)
=  \frac{r_0\,  dr d \phi }{2 \pi \Delta t}
e^{- A^0(\Delta r, \Delta \phi, r_0, \phi_0;\Delta t)}, 
\nonumber\\
\label{G-2d-polar}
\label{G-A-2d-polar}
 \ea
where the time-slice action $A^0(\Delta r, \Delta \phi, r_0, \phi_0;\Delta t)$ is 
\be
A^0(\Delta r, \Delta \phi, r_0, \phi_0;\Delta t) 
= \frac{ \left(\Delta r - \frac{\Delta t}{2 r_0} \right)^2 
 + r_0^2 \Delta \phi^2 } {2 \Delta t} . 
\label{2dfpa}
\ee

Note that Eq.~(\ref{G-2d-polar}) looks very different from Eq.~(\ref{discrete-polar}). Eq.~(\ref{G-2d-polar}) is Gaussian in $\Delta r$ and $\Delta \phi$, whereas by contrast, Eq.~(\ref{discrete-polar}) is clearly not Gaussian.  However Lemmas 1 and 2 guarantee that these two distributions are equivalent in the sense that they share the same moments up to order $\Delta t$, which are given by
\ba
\langle 1 \rangle &=& 1, 
\nonumber\\
\langle \Delta r \rangle &=& \frac{\Delta t}{2 r_0}, 
\quad \langle \Delta \phi \rangle = 0,
 \label{i-ave-sigma-polar}\\
\sigma^2_{\Delta r} &=& \Delta t, \quad
\sigma^2_{\Delta \phi} = \frac{\Delta t}{r_0^2}. \nonumber
\ea
The most salient feature of Eqs.~(\ref{i-ave-sigma-polar}) is that the average of $\Delta r$ is non-vanishing, and inversely proportional to $r_0$ and hence diminishes with increasing $r_0$.    A non-vanishing average of $\Delta r$ is an inevitable consequence of the curvilinear nature of the polar coordinates.  To see this, consider a particle starting from $(r_0, \phi_0) $, and diffuse isotropically.  As illustrated in Fig.~\ref{fig:Spurious-drift}, the probability of $r$ increasing is larger that of $r$ decreasing, simply because there are more space with larger radius.  

\begin{figure}[t]
	\centering
	\includegraphics[width=1.9in]{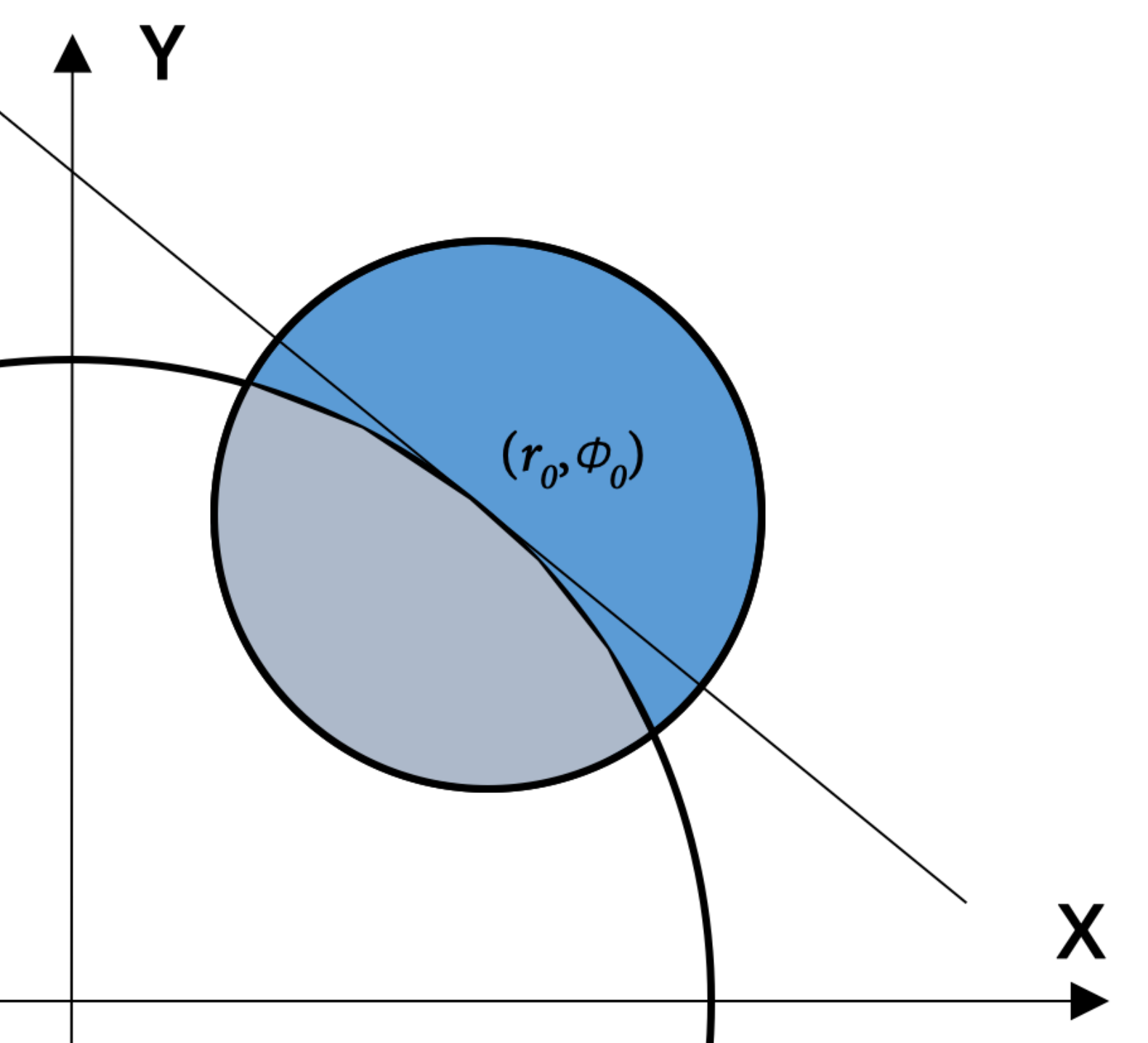}
	\caption{Diffusion in plane with polar coordinates:  Because of the curved nature of the coordinate lines, there is more space with larger radius than with smaller radius.  Hence the probability of $dr>0$ is larger than that of $dr <0$.  } 
	\label{fig:Spurious-drift}
\vspace{-3mm}
\end{figure}

The Hamiltonian Eq.~(\ref{H-2d-polar}) can also be understood as negative of Fokker-Planck operator Eq.~(\ref{FPO-def}) with $U = 0$ and $L^{ij} = B^{ij}= \frac{1}{2} g^{ij}$.  We can then calculate  TSGF using Eqs.~(\ref{trans-amp-classical}), and obtain the same result  Eq.~(\ref{G-A-2d-polar}).  

Finally, using Lemma 3 we also have the  $\alpha$-representation of TSGF:
\ba
&&dv(r, \phi) G( r, \phi | r_0, \phi_0 ; \Delta )
\label{G-S-2d-polar}
 \\
&=& \frac{  r_{\alpha} dr d\phi  }{2 \pi \Delta t} 
 \exp{ \! \left[  - \frac{ ( \Delta r - \frac{ \Delta t}{ 2    r_{\alpha}} )^2 
+   r_{\alpha}^2 \Delta \phi ^2}{2 \Delta t} 
+ \frac{\alpha \Delta t}{2  r_{\alpha}^2}\right]  }. 
\nonumber
\ea
where $  r_{\alpha}= \alpha r + (1- \alpha) r_0$. Compared to Eq.~(\ref{G-A-2d-polar}), the action in Eq.~(\ref{G-S-2d-polar}) has an additional term proportional to $\Delta t$ which is needed to ensure normalization. The spurious drift is independent of $\alpha$ since $\partial_i g^{ij} = 0$ in this case. 



\vspace{-3mm}
\subsubsection{Free particle on a unit sphere}
\vspace{-3mm}

Consider a quantum particle moving on a 2-sphere with unit radius. The Hamiltonian is
\ba
\hat H = - \frac{1}{2 \sin \theta} \partial_\theta \sin \theta \partial_\theta
- \frac{ 1 }{2 \sin^2 \theta} \partial^2_\phi,
\ea
which is in the form of Eq.~(\ref{Hcurved}) with $\xv = (\theta, \phi)$, $g(\xv) = \sin^2 \theta$, and 
\be
g_{ij} = \begin{pmatrix} 
1 & 0\\ 0 & \sin^2 \theta \end{pmatrix}, \quad
g^{ij} = \begin{pmatrix} 
1 & 0\\ 0 &\sin^{-2} \theta  \end{pmatrix}.  
\ee
The TSGF can then be obtained using Eqs.~(\ref{trans-amp-quantum}):
\ba
&& G( \theta, \phi |  \theta_0, \phi_0; \Delta t) dv(\theta, \phi)
\label{G-A-2d-sphere}
\\
&=& \frac{\sin \theta_0  d \theta d \phi }{2 \pi \Delta t}
 e^{-   [ {\left(\Delta \theta -\cot (\theta_0) \Delta t  /2 \right)^2 
 + \sin^2 \theta_0\,  \Delta \phi^2 }]/{2 \Delta t}  }, 
\nonumber
\ea
where $dv(\theta, \phi) = \sin \theta d \theta d \phi$.  
For small $\theta$, these results reduce to Eq.~(\ref{G-A-2d-polar}) with $\theta \rightarrow r$, as it should be. 


\section{Path-integral from Langevin equation}

It is well known that classical Markov processes can  be represented by either a Fokker-Planck equation, or a Langevin equation, or a path-integral representation.  It is also known that in the presence of multiplicative noises, there are infinite versions of equivalent Langevin equations, each corresponding to a particular scheme of stochastic integration.  The transformations between different versions of Langevin theories, as well as between Langevin theory and Fokker-Planck theories are discussed in textbooks, see for example Refs.~\cite{Gardiner-book,van-Kampen-stochastic}.  In the preceding section, we have established the correspondence between the Fokker-Planck theory and the $\alpha$-representation of time-slicing path integral, i.e., Eq.~(\ref{trans-amp-classical}).  In this section, we will show that these representations  can be obtained from the Ito-Langevin dynamics in a remarkably simple way.  This simplicity is a direct consequence of the linear relation between $d \xv$ and Wiener noises $d W_{\mu}$, which is not shared by the Stratonovich-Langevin theory, or other types of Langevin theory where the relation between $d\xv$ and noises is nonlinear. Nonetheless, we will also establish the connection between  time-slicing path integral and $\bar \alpha$-Langevin dynamics (for definition see Eq.~(\ref{alpha-langevin})), and find an equivalent family of path integral actions parameterized by two parameters $\alpha, \bar \alpha$.    As we will see, the resulting action is much more complicated.  


\subsection{ Path-integral from Ito-Langevin equation}
\label{sec:Ito-Langevin-1}
\vspace{-2mm}
We start with the Ito-Langevin equation
\ba
dx^i = F^i (\xv,t) dt + b^{i\mu} (\xv)dW_\mu(t)
\label{LE}
\ea
where $d \xv \equiv \xv(t+ dt) - \xv (t)$ is the infinitesimal evolution of $\xv$ during time step $dt$, and  $dW_\mu(t) , \mu = 1,2,3 \ldots m$ are $m$-dimensional Wiener noises, which are Gaussian and white, acting on $\xv$ during the time interval $(t, t + dt)$:
\be
\langle dW_\mu(t) dW_\nu(t) \rangle = \delta_{\mu\nu} dt.  
\label{Ito-rule}
\ee 
The product $b^{i\mu} (\xv)dW_\mu(t)$ in Eq.~(\ref{LE}) is defined in Ito's sense~\cite{Gardiner-book}, which means that $F^i (\xv,t)$ and $b^{i\mu} (\xv)$  in Eq.~(\ref{LE})  are evaluated at $\xv(t)$.  Hence Eq.~(\ref{LE}) defines a linear relation between $d \xv(t)$ and $dW_\mu(t)$, and $d \xv(t)$ is also Gaussian, whose average and variance can be directly obtained from Eq.~(\ref{LE}):
\begin{subequations}
\label{ave-variance-dx-Ito}
\ba
\langle dx^i \rangle &=& F^i(\xv,t)dt, \\
\Big\langle \big(dx^i - F^i  dt\big)
\big (dx^j -F^i  dt \big) \Big\rangle &=& 2B^{ij}(\xv) dt.  
\ea
\end{subequations}
The symmetric matrix $B^{ij}(\xv)$, assumed non-singular, is related to the noise amplitudes $b^{i \mu}$ in Eq.~(\ref{LE}) via
\ba
B^{ij}(\xv) \equiv  \frac{1}{2} \sum_{\mu} b^{i\mu}(\xv) b^{ju}(\xv)= B^{ji}(\xv). 
\label{B^ij-b^imu}
\ea

The Ito-Langevin equation (\ref{LE}) is mathematically equivalent to the Fokker-Planck equation:
\ba
\partial_t  p =  -   \frac{1}{\sqrt{g }} 
\partial_i \big( F^i  \sqrt{g } \, p \big) 
+  \frac{1}{\sqrt{g }}  \partial_i \partial_j 
\big(B^{ij}\sqrt{g }  \, p \big),
\quad \label{LFP2}
\ea
where $p(\xv,t) \sqrt{g(\xv)} d^d x$ is the  differential probability. In two preceding publications~\cite{Ding2020, Ding2021}, we and collaborator  formulated a covariant Ito-Langevin theory, where $F^i$ in Ito-Langevin equation (\ref{LE}) is parameterized as
\ba
F^i  &=& \frac{1}{\sqrt{g}} \left( \partial_j 
\sqrt{g}L^{ij} \right) - L^{ij} \partial_j U,
\label{F-L-U}
\ea 
\\where $U$ is related to the steady state probability density function (pdf) via $U(\xv) = - \log p^{\rm SS}(\xv)$, and $L^{ij}$ can be decomposed a symmetric part $B^{ij}$ and an antisymmetric part $Q^{ij}$, the latter being related to the steady state current via: $J_{\rm SS}^i = \partial_j \left( Q^{ij} e^{-U} \right)$.   The Ito-Langevin equation (\ref{LE}) and Fokker-Planck equation (\ref{LFP2}) then can be rewritten into the covariant forms, Eqs.~(\ref{Langevin-0}) and (\ref{FPE-0}).   The precise meaning of covariance is discussed in Refs.~\cite{Ding2020,Ding2021}, and will be discussed in more detailed in Sec.~\ref{sec:covariance}.  


Since $d\xv$ as determined from Ito-Langevin dynamics is Gaussian, we can directly write down its probability distribution using its first and second order moments, given in Eqs.~(\ref{ave-variance-dx-Ito}).  Let $\xv_1 \equiv \xv + d \xv $, and $d\mu(\xv_1) = d^d x_1$  the infinitesimal volume element at $\xv_1$ and $dv(\xv_1) = \sqrt{g(\xv_1) } d\mu(\xv_1)$, we have
\ba 
&& dv(\xv_1) \, G( \xv_1 | \xv ; dt) = 
\label{PIIL}\\
&& \frac{d\mu(\xv_1) \, e^ { - 
(d x^i - F^i(\xv,t) \, dt ) \frac{B^{-1}_{ij}(\xv) }{ 4 dt} 
 (d x^j - F^j(\xv,t) \, dt ) } }
{  \sqrt{ (4 \pi)^d \det B^{ij}(\xv)  dt}}. 
\nonumber
\ea
Note that Eq.~(\ref{PIIL}) refers to transition from $\xv$ to $\xv_1 = \xv + d \xv$ during the time interval $dt$, whereas in Eq.~(\ref{trans-amp-classical}), the transition is from $\xv_0 $ to $\xv_0 + \Delta \xv = \xv$ during the time interval $\Delta t$.  With the correspondence of notations $(dt, d\xv, \xv, \xv_1) \leftrightarrow (\Delta t, \Delta \xv, \xv_0, \xv)$, and $F^i$ given by Eqs.~(\ref{F-L-U}) and (\ref{PIIL}) indeed reduces to Eqs.~(\ref{trans-amp-classical-1}) and (\ref{trans-amp-classical}).   

\subsection{Path-integral for Stratonovich-Langevin  and $\bar \alpha$-Langevin}
Many physicists prefer Stratonovich version of Langevin equation:
\begin{subequations}
\label{Slangevin-combo}
\ba
dx^i = F^i_S(\xv,t) dt + b^{i\mu}(\xv,t) \circ dW_\mu(t),
\label{Slangevin}
\ea
where the product  $b^{i\mu}(\xv,t) \circ dW_\mu(t)$ is defined in Stratonovich's sense:
\be
b^{i\mu}(\xv,t) \circ dW_\mu(t) \equiv b^{i\mu}(\xv + d \xv/2,t) \,  dW_\mu(t), 
\label{b-W-Strato}
\ee
\end{subequations}
which means that the noise amplitudes are evaluated as the midpoint $\xv + d \xv/2$.  Because of the hidden  dependence of Eq.~(\ref{b-W-Strato}) on $d\xv$,  the relation between $d\xv$ and the Wiener noises $dW_\mu(t)$, as defined by Eq.~(\ref{Slangevin}) is more complicated.  This leads to substantial complexity in the calculation of pdf for $d\xv$.   However, using Ito's formula, one can easily prove~\cite{Gardiner-book} that the Stratonovich-Langevin equation (\ref{Slangevin}) is equivalent to the Ito-Langevin equation (\ref{LE}) with the following correspondence:
\ba 
F^i(\xv,t)  =  F^i_S(\xv,t) + \frac{1}{2} \,b^{j \mu}(\xv,t) \partial_j b^{i\mu} (\xv,t) .  
\label{S-Langevin-F}
\ea

More generally there is a continuous family of representations which is parameterized by  $\bar \alpha \in [0,1]$:
\begin{subequations}
\be
dx^i = F^i_{\bar \alpha} (\xv,t) dt 
+ b^{i\mu}(\xv,t) \otimes_{\bar \alpha} dW_\mu(t),
\label{alpha-langevin}
\ee
where the product $b^{i\mu}(\xv,t) \otimes_{\bar \alpha} dW_\mu(t)$ is defined as
\be
b^{i\mu}(\xv,t) \otimes_{\bar \alpha} dW_\mu(t) \equiv
 b^{i\mu}(\xv + {\bar \alpha} \,d \xv, t) \,  dW_\mu(t),  
\label{b-W-alpha}
\ee
which means that the noise amplitudes $b^{i \mu}$ are evaluated at an intermediate point $\xv_{\bar \alpha} \equiv \xv + {\bar \alpha} \,d \xv $.    We shall call Eq.~(\ref{alpha-langevin}) an {\em $\bar \alpha$-Langevin equation}.  Using Ito's formula, one can easily prove that the $\bar \alpha$-Langevin equation (\ref{alpha-langevin}) is equivalent to Ito-Langevin equation (\ref{LE}) with the following correspondence:
\ba 
F^i(\xv,t)  =  F^i_{\bar \alpha} (\xv,t) +\bar \alpha \,
 b^{j \mu}(\xv,t) \partial_j b^{i\mu} (\xv,t) .  
\label{alpha-Langevin-F}
\ea
\label{alpha-langevin-combo}
\end{subequations}
It is important to note that the parameter $\bar \alpha$ introduced in Langevin equation (\ref{b-W-alpha}) is independent of the parameter $\alpha$ we introduce earlier in path integral action in Sec.~\ref{sec:alpha-rep}.  Whilst in some previous works people often identified these two parameters, they is {\it a priori} no  reason for them to be the same.  

We can use Eq.~(\ref{S-Langevin-F}) to replace $F^i(\xv,t) $ in Eq.~(\ref{PIIL}) in terms of $F^i_S(\xv,t)$, and obtain an equivalent expression for the TSGF: 
\begin{widetext}
\vspace{-4mm}
\ba
dv(\xv_1)  \, G( \xv_1 | \xv; dt)
&=&  \frac{d\mu(\xv_1)   \exp \left\{- \left( dx ^i - F_S^i dt 
 - \frac{1}{2}b^{k\mu }\partial_k b^{i\mu} dt \right)
 \frac{B^{-1}_{ij} }{4 dt}  
\left(dx^j - F_S^j dt -\frac{1}{2}b^{l \nu}\partial_l b^{j \nu} dt \right)
\right\}} { \sqrt{ (4 \pi)^d \det B^{ij} } }  ,
\ea
where the functions $F_S, B, B^{-1}$ and $b$ are evaluated at $\xv$.  Further applying  Lemma 3, we obtain an equivalent representation for  $G( \xv_1 | \xv; dt)$ where  all functions are evaluated at the  Stratonovich point  $\xv_{1/2} = \xv + d \xv/2$: 
\ba
&& dv(\xv_1) \, G( \xv_1 | \xv; dt)
=  \frac{d\mu(\xv_1)  }
{ \sqrt{ (4 \pi) ^d \det B^{ij}(\xv_{1/2}) } }  \exp \bigg\{
\nonumber\\
&-  & \Big(dx ^i   - F_S^i(\xv_{1/2}) dt
 + \frac{1}{2}  b^{i\mu}(\xv_{1/2})   \partial_k   b^{k \mu}(\xv_{1/2}) dt  \Big)
\frac{B^{-1}_{ij} (\xv_{1/2}) }{4 dt}  
\Big(dx ^j  - F_S^j(\xv_{1/2}) dt
 + \frac{1}{2}  b^{i\mu}(\xv_{1/2})  \partial_k b^{k \mu}(\xv_{1/2})  dt \Big) 
 \nonumber \\
 &-& \frac{dt}{2} \partial_i F_S^i 
+ \frac{dt}{8} \left( (\partial_i  b^{i\mu}) ( \partial_j b^{j \mu} )
- (\partial_i  b^{j \mu}) ( \partial_j b^{i\mu} ) \right) \bigg\}. 
\label{Stralpha}
\ea
Recall that it does not matter whether we evaluate the last two terms in Eq.~(\ref{Stralpha}) (both linear in $dt$) at $\xv$ or at $\xv_{1/2}$.

%

More generally, we may use the $\bar \alpha$-Langevin equation (\ref{alpha-langevin-combo}) and evaluate all functions at $\xv_{\alpha}= \xv+ \alpha d\xv$. The resulting TSGF is even more complicated:
\ba
dv(\xv_1) \, G( \xv_1 | \xv; dt)
& = & \frac{d\mu(\xv_1)  }
{ \left (  4 \pi  \right )^ {d/2} \sqrt{\det B^{ij} (\xv_{\alpha})}}
\exp \bigg\{
\nonumber\\ 
- \Big(dx ^i   - F_{\bar \alpha}^i dt 
&-& \bar \alpha \, b^{k\mu} \partial_k b^{i\mu}dt
 + 2 \alpha \, \partial_k B^{ik} dt  \Big)_{\! \alpha}
\frac{B^{-1}_{ij}  (\xv_{\alpha})}{4 \, dt}  
\Big(dx ^j  - F_{\bar \alpha}^j dt
 - \bar \alpha \, b^{l\nu} \partial_l b^{j\nu}dt
 +2 \alpha \, \partial_l B^{jl} dt \Big)_{\! \alpha} 
\nonumber\\ 
- \alpha \, \partial_i F_{\bar \alpha}^i  \, dt
  &-& \alpha \bar \alpha\, \partial_i ( b^{k\mu} \partial_k b^{i\mu}) \, dt
   + \alpha^2 \partial_i\partial_j B^{ij} \, dt  \bigg\}. 
\label{alphaLpi}
\ea
\end{widetext}
 The spurious drift is now given by  $( \bar \alpha \, b^{k\mu} \partial_k b^{i\mu} -  2 \alpha \, \partial_k B^{ik} ) dt  $, which depends both on $\alpha$ and on $\bar \alpha$.  It can be seen that for a general multi-dimensional model, there is no way to cancel the spurious drift  by tuning of parameters $\alpha, \bar \alpha$.  For $\alpha = \bar \alpha = 1/2$, Eq.~(\ref{alphaLpi}) reduces to Eq.~(\ref{Stralpha}).  For $\alpha = \bar \alpha = 0$, Eq.~(\ref{alphaLpi}) reduces to Eq.~(\ref{PIIL}), which is  the simplest of all representations.  The case $\alpha = \bar \alpha$ is in accordance with the result given by Langouche {\it et. al.} \cite{Langouche-book,Langouche1978-1}. For 1d case and $\alpha = \bar \alpha$, our result further reduces to that of Lau and Lubensky~\cite{Lau2007}. 

\subsection{Edwards-Gulyaev revisited}
Let us revisit the problem of Edwards and Gulyaev using Langevin theory.  The Langevin equations of a free particle in Cartesian coordinates are
\ba
dx = dW_1(t) ; \quad dy =  dW_2(t). 
\label{dx-dy-1}
\ea
The transition probability is just 
\ba
&& dxdy\,G ( x_1,y_1|x,y; dt ) 
= \frac{dxdy}{2\pi} \exp{ - \frac{ dx^2 + dy^2}{2 dt}},
\nonumber\\
\label{tpcc-2}
\ea
where $dx = x_1 - x, \quad dy = y_1 - y$. 
Now transforming to  polar coordinate using Eq.~(\ref{polar-transform}) and invoking the property of the Winer process $dW_1(t)^2 = dW_2(t)^2 = dt, \quad dW_1(t) dW_2(t) = 0$, we obtain
\begin{subequations}
\ba
dr & =& \cos \phi \, dx + \sin\phi \, dy 
 +\frac{1}{2} \frac{ \partial^2 r}{ \partial x^2} \, dx^2
+ \frac{1}{2} \frac{ \partial^2 r}{ \partial y^2}\,  dy^2
\nonumber \\ 
& = & \cos \phi \, dx + \sin \phi \, dy + \frac{dt}{2r};
\\
d\phi &=& - \frac{\sin \phi}{r} dx + \frac{\cos\phi}{r} dy + 
\frac{ \partial^2 \phi}{ \partial x^2} dx^2
+  \frac{ \partial^2 \phi}{ \partial y^2} dy^2
\nonumber\\
& = & - \frac{\sin \phi}{r} dx + \frac{\cos\phi}{r} dy.
\ea
\end{subequations}
Using these to express $dx, dy$ in terms of $dr, d \phi$ in Eqs.~(\ref{dx-dy-1}), we obtain the Ito-Langevin equations in polar coordinates:
%
\begin{subequations}
\ba
dr & = & \frac{1}{2r} dt + b^{1 m} dW_m; \\
d\phi & = & b^{2 m} dW_m.
\label{ILP}
\ea
\label{ILP-0}
\end{subequations}
where the matrix $b^{i m}$ is given by
\ba
\left( b^{i m} \right) =  \begin{pmatrix}
\cos \phi &  \sin \phi \\
-\sin \phi/r & \cos \phi/r 
\end{pmatrix},
\label{jacobianpolar}
\ea
from which we can construct $B^{ij}$ using Eq.~(\ref{B^ij-b^imu}).  

So we can write the transition probability as
\ba
&& dv(r_1,\phi_1) G(  r_1, \phi_1 | r , \phi,dt )
\label{tppc-3}\\ 
& = & \frac{r d\mu(r_1, \phi_1) }{ 2\pi dt} \exp{ -\bigg(  \frac{ ( dr - dt/(2r))^2}{2 dt}
+ \frac{ r^2 d\phi^2}{2dt}\bigg)} , \nonumber
\ea
where $dv(r_1,\phi_1) = r_1 dr_1 d \phi_1 $ and $d\mu = dr_1d\phi_1$.
This result is identical to Eqs.~(\ref{G-A-2d-polar}), under the correspondence of notations: $(dr_1, d\phi_1, dt, r, \phi) \leftrightarrow 
(\Delta r, \Delta  \phi, \Delta t, r_0, \phi_0)$. 
 
The Ito-Langevin equations (\ref{ILP-0}) can also be transformed into Stratonovich-Langevin form:
\begin{subequations}
\ba
dr & = & \frac{1}{2r} dt + b^{1 m}\circ dW_m; \\
d\phi & = & b^{2 m} \circ dW_m.
\label{SLP}
\ea
\end{subequations}
which is formally identical to the Ito form (\ref{ILP-0}), except that the Ito products are replaced by the Stratonovich products.   The path-integral representation with $\alpha = 1/2$ can be obtained using Eq.~(\ref{Stralpha}):
\ba
&&dv(r_1,\phi_1)  G( r_1, \phi_1 | r, \phi ; dt )
\label{G-S-2d-polar-1}
 \\
&=& \!\! \frac{ r_{1/2}d\mu(r_1,\phi_1) }{2 \pi dt} 
 \exp{  \!\! \left[  - \frac{ ( dr - \frac{ d t}{ 2    r_{1/2}} )^2 
+   r_{1/2}^2 d \phi ^2}{2 d t} 
+ \frac{dt}{4  r_{1/2}^2}\right]  },
\nonumber
\ea
which coincides with Eq.~(\ref{G-S-2d-polar}) under the correspondence: $(dr_1, d\phi_1, dt, r, \phi) \leftrightarrow (\Delta r, \Delta  \phi, \Delta t, r_0, \phi_0)$, and $\alpha \rightarrow 1/2$. 

\section{Covariance}
\label{sec:covariance}
The principle of covariance dictates that all basic equations of a physics theory must be represented in tensor forms, which transform according to tensor algebra under nonlinear transformation of variables (NTV).  As such, these equations have the same forms in different coordinate systems, and their validity is independent of choice of coordinate systems.  The principle of covariance has served as a cornerstone for the major parts of modern theoretical physics, including general relativity and gauge theories of elementary interactions.  The fundamental assumption underlying this principle is that laws of physics are objective, whereas choices of coordinate system are subjective.  Change of coordinate system does not change physical laws, but only leads to equivalent representations of the same laws.  


In the setting of time-slicing path integral, the issue of covariance turns out to be more subtle than that in general relativity.  This is because the objects $\Delta \xv$, which appears ubiquitously in TSGF, are not infinitesimal vectors in conventional sense.   It is important to note that even though a large number of works~\cite{Graham1977,Graham(19772), Cugliandolo2019,Weiss(1978),Kerler1978,Hirshfeld-1978,McLaughlin1971,Langouche-book,Langouche1978,Durr1978}  were published on the topic of path integral in curved space, most of these works do not address explicitly how action transform under general NTV.  Edwards and Gulyaev~\cite{Edwards1964} argued that the usual chain rule of calculus is not applicable in coordinate transformation of path integral, but did not supply detail. Deininghaus and Graham~\cite{Deininghaus1979} developed path integral in curved space using normal coordinate systems, which was also discussed and developed by Langouche {\it et. al.}~\cite{Langouche-book}.  More recently, Cugliandolo {\it et. al.} ~\cite{Cugliandolo2017,Cugliandolo2018,Cugliandolo2019} made one dimensional path integral covariant under usual calculus rules by adjusting the interpolation parameter $\alpha$.  It remains to be shown whether this approach works for higher dimensions.  
 


We first note that we have formulated both the quantum Hamiltonian Eq.~(\ref{Hcurved}) and the classical Fokker-Planck operator Eq.~(\ref{FPO-def}) in terms of tensors and transform as scalars under NTV.    For the quantum case, the tensor objects are covariant metric tensor  $g_{ij}$ and scalar potential $V$.  For the classical case, the tensor objects are  the contra-variant tensor $L^{ij}$, the covariant metric tensor $g_{ij}$, and the scalar $U(\xv)$.  Our path integral representations for TSGT, Eqs.~(\ref{trans-amp-quantum}) for the quantum case and Eqs.~(\ref{trans-amp-classical-neq-0}) for the classical case, are also formulated in terms of these tensor objects.  When one makes a NTV, these tensor objects transform in the following way: 
\begin{subequations} 
\label{transform-all-new}
\ba
p (\xv) &\rightarrow& p'(\xv') = p(\xv),
\label{covariant-rules-1-new}\\
V (\xv )  &\rightarrow& V'(\xv') = V (\xv) , 
\label{covariant-rules-V-new}\\
U (\xv )  &\rightarrow& U'(\xv') = U (\xv) , 
\label{covariant-rules-U-new} \\
g_{ij}(\xv )  &\rightarrow& g'_{ab} (\xv' )  
= \frac{\partial x^i}{\partial x'^a} g_{ij}(\xv )  
\frac{\partial x^j}{\partial x'^b},
\label{covariant-rules-g-new}\\
b^{i \mu}(\xv) &\rightarrow& b'^{a \mu}(\xv') 
= \frac{\partial x'^a}{\partial x^i}
b^{i \mu}(\xv),
\label{covariant-rules-b-new}\\
L^{ij}(\xv) &\rightarrow& L'^{ab}(\xv') 
= \frac{\partial x'^a}{\partial x^i} L^{ij} (\xv)
\frac{\partial x'^b}{\partial x^j}. 
\label{covariant-rules-3-new}
\ea
\end{subequations} 
We can now use the transformed tensors listed in Eqs.~(\ref{transform-all-new}) to construct $\alpha$-representation of TSGF in new coordinate systems.  For example, the $\alpha = 0$ representation of quantum TSGF in the new coordinate system is given by
\vspace{-2mm}
\begin{widetext}
\vspace{-4mm}
\begin{subequations}
\label{trans-amp-quantum-new}
\ba
dv(\xv') \, G'(\xv' | \xv'_0; \Delta t ) = 
\frac{\sqrt{g'(\xv'_0)}\, d\mu(\xv') }{(2 \pi \hbar \Delta t)^{d/2}}
e^{- A'^0 (\xv', \xv'_0; \Delta t)},  
\label{trans-amp-quantum-1-new}
\ea
\vspace{-1mm}
\ba
  A'^0 (\xv', \xv'_0; \Delta t) &=&  \left[  \Delta x'^a 
-  \frac{\hbar \Delta t}{2 } 
\left(  \frac{1}{\sqrt{g'}}\, \partial'_c  \sqrt{g'}  g'^{ca} \right)_{\!\! 0} \, \right] 
\frac{g'_{ab}(\xv'_0)}{2 \hbar \Delta t}
\left[   \Delta x'^b -   \frac{\hbar \Delta t}{2} 
\left( \frac{1}{\sqrt{g'}}\,\partial'_d \sqrt{g'} \,  g'^{db} \right)_{\!\! 0}  \, \right] 
+ \frac{\Delta t}{\hbar} V'(\xv'_0),
\nonumber\\
&=& \left[   \Delta x'^a 
+  \frac{\hbar \Delta t}{2 }  g'^{cd}(\xv'_0) \Gamma'^{a}_{cd} (\xv'_0)  \right] 
\frac{g'_{ab}(\xv'_0) }{2 \hbar \Delta t}
\left[  \Delta x'^b + \frac{\hbar \Delta t}{2} 
  g'^{ef}(\xv'_0) \Gamma'^{j}_{ef}(\xv'_0)   \right] 
+ \frac{\Delta t}{\hbar} V'(\xv'_0),
\label{trans-amp-quantum-2-1-new}
\ea
\end{subequations}
where $\Delta x'^a = x'^a - x_0'^a$.  Obviously Eqs.~(\ref{trans-amp-quantum-new}) have the same forms as Eqs.~(\ref{trans-amp-quantum}), but with all tensors and coordinates replaced by the transformed versions.   Similar constructions can be made for $\alpha \neq 0$ representations of quantum TSGF, and also for the classical TSGF. 
\end{widetext}


But one may also attempt to apply NTV directly to the action $A^\alpha (\xv, \xv_0; \Delta t) $ in Eq.~(\ref{qmalpha-0}), and hope to get the action $A'^\alpha (\xv', \xv_0'; \Delta t) $.  This  expectation seems rather natural, since it is known that the classical action (\ref{action-curved}) transform as a scalar.  Careful analysis however indicates that the action for time-slicing path integral is not a scalar in usual sense.   In usual calculus, $\Delta \xv$  behaves as a covariant infinitesimal vector: $\Delta x'^a = \left( \partial x'^a /\partial x^i \right) \Delta x^i$.  But this transformation law can not hold in the presence case, because we know that quadratic terms $\Delta \xv \Delta \xv$ are of order $\Delta T$ and hence important.  On the other hand, if we treat $\Delta \xv, \Delta \xv'$ as finite quantities, then $\Delta \xv$ should be treated as a fully nonlinear function of $\Delta \xv'$.  Substituting this function into Eq.~(\ref{qmalpha-0}) we would obtain a representation of  TSGF that is generically outside the domain of $\alpha$-representation.  

The example of Edwards-Gulyaev serves to illustrate these results.  The action in Cartesian coordinates is given by Eq.~(\ref{action-EG-Cartesian}). If $\Delta x, \Delta y$ transform as usual vectors, we would obtain $(\Delta r^2 + r_0^2 \Delta \phi^2)/2 \Delta t$ as the transformed action, which, comparing with the correct result Eq.~(\ref{2dfpa}), misses the spurious drift.  On the other hand, if we treat $\Delta x, \Delta y$ as finite quantities, we would obtain the TSGF as given by Eq.~(\ref{discrete-polar}), which is of course correct.  However the resulting action does not have the form of $\alpha$-representation of TSGF.  

These results may appear very surprising, because in the definition of Green's function, Eq.~(\ref{P-transition-0}),  both the operator $\hat L$ and the Dirac delta function $\ddelta(\xv, \xv_0 ) $ are scalars.  Hence the Green's function and the TSGF must also be scalars. But then how do we understand the transformation rules of time-slice action?  

Recall that the $\alpha$-representations of TSGF are constructed to reproduce all moments of TSGF up to the order of $\Delta t$.  There are infinitely many other representations of TSGF which do not assume the form of Eqs.~(\ref{qmalpha-all}) and (\ref{trans-amp-classical-neq-0}) but yet they are equivalent in the sense of Lemma 1.  To understand the covariance of our $\alpha$-representations, we must confine ourself within the domain of these representations.  Below we will show that for $\alpha = 0$, the time-slicing action transforms as a scalar, if $\Delta \xv$ transform according to Ito's formula.

\vspace{-2mm}
\subsection{ { $\alpha = 0$ Representation of Classical Path Integral}}
Let us first discuss the case of classical stochastic processes.  In two previous works~\cite{Ding2020, Ding2021}, we and collaborator have established the covariance of both Ito-Langevin theory and Fokker-Planck theory.  More specifically, we have shown that under the transformation rules Eqs.~(\ref{transform-all-new}), the Fokker-Planck equation (\ref{FPE-0}), which we rewrite below:
\begin{align}
\partial_t \, p  
=  \frac{1}{\sqrt{g}} \partial_i \sqrt{g} L^{ij} (\partial_j + (\partial_j U)) p
= \hat L_{\rm FP} p ,
\tag{14}
\end{align}
 is transformed into:
\ba
\partial_t p'  &=& \frac{1}{\sqrt{g'}} 
\partial'_a  \sqrt{g'}  L'^{ab} \big( (\partial'_b U') 
 + \partial'_b \big) p' = \hat L_{\rm FP}' p', 
 \quad \quad  \label{FPE-new}
\ea
which has the same form as Eq.~(\ref{FPE-0}).  The Fokker-Planck operator transforms also as a scalar:
\be
\hat L_{\rm FP}'(\xv') =\hat  L_{\rm FP}(\xv).   
\ee
Furthermore, under the  transformation rules Eqs.~(\ref{transform-all-new}) together with  {\em Ito's formula}~\cite{Gardiner-book}:
\ba 
d x'^a 
&=& \frac{\partial x'^a}{ \partial x^i} \, dx^j  
+ B^{ij} \frac{\partial^2 x'^a}{\partial x^i \partial x^j} dt,
\label{transform-dx-Ito}
\ea
the Ito-Langevin equation (\ref{Langevin-0}), which we rewrite below
\begin{align}
dx^i + \left(  L^{ij} \partial_j U - \frac{1}{\sqrt{g}} 
\partial_i \sqrt{g} L^{ij} \right) dt
=  b^{i\mu} dW_\mu(t),
\tag{13}
\end{align} 
 is  transformed into:
\ba
&& dx'^a + \left( L'^{ab} \partial'_b U'  
 - \frac{1}{\sqrt{g'}} \partial'_b \sqrt{g'} L'^{ab}  \right)dt
 = b'^{a\alpha} d W^\alpha,
\nonumber\\
  \label{CoLE-prime} 
\ea
which again has the same form as Eq.~(\ref{Langevin-0}).  Ito's formula (\ref{transform-dx-Ito}) was derived by using Eq.~(\ref{Langevin-0}) and the property of Wiener noises (\ref{Ito-rule}), and keeping terms up to $O(dt)$~\cite{Gardiner-book,van-Kampen-stochastic,Jacobs2010-book}. 

Equation (\ref{transform-dx-Ito}) can be understood as a linear inhomogeneous relation between two stochastic variables $d\xv, d\xv'$.  Taking the average, we find 
 \ba
\left\langle  d x'^a \right\rangle &=& 
\frac{\partial x'^a}{\partial x^i}  \left\langle d x^i \right\rangle
+ \frac{\partial^2 x'^a}{\partial x^i \partial x^j}   B^{ij} d t,
\label{ave-Ito-formula-classical}
\ea
But according to the Langevin equations, the averages of $d x^i$ and $d x'^a$ are respectively:
\begin{subequations}
\ba
\langle d x^i \rangle &=&  - L^{ij} \partial_j U dt 
+ \frac{1}{\sqrt{g}} \partial_j \sqrt{g} L^{ij} dt, \\
\langle d x'^a \rangle &=&  - L'^{ab} \partial'_b U' dt \
+ \frac{1}{\sqrt{g'}} \partial'_b \sqrt{g'} L'^{ab} dt. 
\ea
\end{subequations}
Substituting these back into Eq.~(\ref{ave-Ito-formula-classical}), we obtain:
\ba
&& - L'^{ab} \partial'_b U' 
+ \frac{1}{\sqrt{g'}} \partial'_b \sqrt{g'} L'^{ab} dt
\label{transform-F}
\label{transform-F-Ito} \\
&=& \!\! \frac{\partial x'^a}{\partial x^i} \left(
 - L^{ij} \partial_j U  + \frac{1}{\sqrt{g}} \partial_j \sqrt{g} L^{ij} 
 \right) \!
+ \! \frac{\partial^2 x'^a}{\partial x^i \partial x^j}   B^{ij} . 
\nonumber
\ea
According to Eqs.~(\ref{transform-all-new}), $L^{ij} \partial_j U$ transforms as a vector, hence we further obtain:
\ba
&&  \frac{1}{\sqrt{g'}} \partial'_b \sqrt{g'} L'^{ab} 
= \!\! \frac{\partial x'^a}{\partial x^i} 
\left( \frac{1}{\sqrt{g}} \partial_j \sqrt{g} L^{ij}  \right) \!
+ \! \frac{\partial^2 x'^a}{\partial x^i \partial x^j}   B^{ij}. 
\nonumber\\
\label{transform-F-1}
\label{transform-F-Ito-1} 
\quad 
\ea
This relation actually can be directly obtained by using Eqs.~(\ref{transform-all-new}), together with the identity (proved in the appendix of Ref.~\cite{Ding2020}):
\be
\frac{\partial}{\partial x'^a} 
\left( \frac{\partial x'^a}{\partial x^i} J \right) = 0, 
\ee
where $J = \det \left( \frac{\partial x'^a}{\partial x^i} \right)$ is the Jacobian,  

Combining  Eqs.~(\ref{transform-dx-Ito}) with (\ref{transform-F-Ito}), we see that even though neither $dx^i $ nor $ \left(  L^{ij} \partial_j U - \frac{1}{\sqrt{g}} 
\partial_i \sqrt{g} L^{ij} \right) dt$  transforms as a vector, the linear combination $dx^i + \left(  L^{ij} \partial_j U - \frac{1}{\sqrt{g}} \partial_i \sqrt{g} L^{ij} \right) dt$, which constitutes the LHS of Eq.~(\ref{Langevin-0}),  does transform as a contra-variant vector.  But the same combination also appear in the action given in Eq.~(\ref{trans-amp-classical}) (here we identify $\Delta \xv$ with $d \xv$, and $\Delta t$ with $dt$).  We hence conclude that {\em the time-slice action $A^0( \xv, \xv_0, \Delta t)$ for classical Markov process transforms as a scalar under NTV, if $\Delta \xv$ transform according to Ito's formula (\ref{transform-dx-Ito}):}
\ba 
\Delta x'^a 
&=& \frac{\partial x'^a}{ \partial x^i} \, \Delta x^j  
+ B^{ij} \frac{\partial^2 x'^a}{\partial x^i \partial x^j} \Delta t.
\label{transform-dx-Ito-3}
\ea
Ito-Langevin equation are mostly frequently written in an alternative form of Eq.~(\ref{LE}), where the  systematic force  appear in RHS.  From the perspective of covariance, however, the form Eq.~(\ref{LE}) is less convenient, because neither side of it transforms as a vector.    

We have not found any simple transformation rule for $\Delta \xv$ which makes {the time-slice action $A^{\alpha}( \xv, \xv_0, \Delta t)$ with $\alpha \neq 0$ invariant.  It hence appears that the $\alpha = 0$ representation of time-slicing path integral is special.  

\subsection{{ $\alpha = 0$ Representation of Quantum Path Integral }}
\label{sec:covariance-quantum}
We can discuss the covariance of quantum time-slicing path integral by making analogy with the classical case.  Firstly, we can read off the first moments of $\Delta x'^a$ and $\Delta x^i$ from Eqs.~(\ref{trans-amp-quantum-new}) and (\ref{trans-amp-quantum}):
\begin{subequations}
\label{dx'-dx-ave-quantum}
\ba
\left\langle  \Delta x'^a \right\rangle &=&
 \frac{\hbar \Delta t}{2 m \sqrt{g'}} 
\,\partial'_b \sqrt{g'} 
\,  g'^{ab} ,
\label{ave-Ito-formula-3}\\
\left\langle  \Delta x^i \right\rangle &=&
 \frac{\hbar \Delta t}{2 m \sqrt{g}} 
\, \partial_k  \sqrt{g}  g^{ik}.
\ea
\end{subequations}
Alternatively, we can also treat $\Delta x'^a$ as a function of $\Delta \xv$ and $\xv_0$, expand it in terms of $\Delta \xv$ up to the second order: 
\ba
\Delta x'^a = \frac{\partial x'^a}{\partial x^i} \Delta x^i 
+ \frac{1}{2}\frac{\partial^2 x'^a}{\partial x^i \partial x^j} 
\Delta x^i \Delta x^j + \cdots. 
\label{x'-x-expansion}
\ea
Taking average of this, and using Eqs.~(\ref{ave-moments-dx}) we find 
 \ba
\left\langle  \Delta x'^a \right\rangle \!\! 
 &=&  \frac{\partial x'^a}{\partial x^i}  \left\langle  \Delta x^i \right\rangle
 + \frac{\hbar \Delta t}{2 m }  
 \frac{\partial^2 x'^a}{\partial x^i \partial x^j}  g^{ij} .  
\label{ave-Ito-formula-2}
\ea
This relation indicates that the average of $\Delta \xv$ transforms inhomogeneously under NTV.

Combining Eqs.~(\ref{dx'-dx-ave-quantum}) and (\ref{ave-Ito-formula-2}) we obtain:
\ba
 \frac{1}{\sqrt{g'}}\,\partial'_b \sqrt{g'} \,  g'^{ab} 
= \frac{\partial x'^a}{\partial x^i} 
 \frac{1}{\sqrt{g}}\, \partial_k  \sqrt{g}  g^{ik}
+ \frac{\partial^2 x'^a}{\partial x^i \partial x^j}  g^{ij}, \nonumber\\
\label{Drift-transform-quantum}
\ea
which can actually be proved independently, just like Eq.~(\ref{transform-F-Ito-1}).  
%


%
%

Now comes the most interesting observation.  If we impose the following transformation law on $\Delta \xv$:
\be
\Delta x'^a = \frac{\partial x'^a}{\partial x^i} \Delta x^i 
+ \frac{\hbar}{2 m } \, g^{ij} 
\frac{\partial^2 x'^a}
{\partial x^i \partial x^j}  \Delta t,
\label{quantum-Ito}
\ee
the linear combination 
$$ \Delta x^i -  \frac{\hbar \Delta t}{2 m \sqrt g } 
\, \partial_k  \sqrt{g}  g^{ki} $$ 
transforms as a contra-variant vector, and consequently the $\alpha = 0$ time-slice action Eq.~(\ref{trans-amp-quantum-2}) transforms as a scalar, i.e., it remains invariant under NTV.  We call Eq.~(\ref{quantum-Ito}) the {\em quantum Ito's formula} because it involves the Planck constant $\hbar$, and is closely related to Ito's formula (\ref{transform-dx-Ito-3}) in the classical case.

\vspace{-2mm}
\section{Conclusion and Outlook}
Let us summarize the main results we have achieved in this work: (1) With Lemma 1 we have established a rigorous criterion for equivalence between different representations of  time-slicing path integrals.  (2) With Lemma 2 we have demonstrated the existence of a Gaussian representation for the time-slice Green's function.  (3) With Lemma 3 we  constructed a continuous family of equivalent time-slicing path-integral actions, parameterized by an interpolation parameter $\alpha \in [0,1]$.  (4) We have established the connection between Langevin dynamics and time-slicing path integral, and have clarified the origin and transformation rules of spurious drift.  (5) We have explained why in general path integral actions do not transform as scalars under usual rules of tensor calculus.  We have also explicitly demonstrated that the $\alpha = 0$ representation of path integral action transforms as a scalar if $\Delta \xv$ transform according to Ito's formula, while all other ingredients transform according to the usual rules of tensor calculus.  Our results resolve several major confusions regarding time-slicing path integral in curved space, and provide a rigorous, practical, and manifestly covariant calculation scheme.  

In the future, we shall apply the path-integral method developed here to study concrete  quantum and classical stochastic dynamics in curved space.   We shall also try to develop a similar method for field theory in curved space, as well as for non-abelian gauge theory and nonlinear sigma field theory.   
 
\section{acknowledgement}

The authors acknowledge support from NSFC via grant 11674217, Shanghai Municipal Science and Technology Major Project (Grant No.2019SHZDZX01), as well as additional support from a Shanghai Talent Program.

\appendix

\bibliographystyle{quantum}

\begin{thebibliography}{99}





\bibitem{Dirac(1947)}
Dirac, Paul Adrien Maurice. The principles of quantum mechanics. No. 27. Oxford university press, 1981.

\bibitem{Feynman(1948)} 
Feynman, R. P. Space-Time Approach to Non-Relativistic Quantum Mechanics.
\href{https://doi.org/10.1103/RevModPhys.20.367}
{Reviews of Modern Physics \textbf{20}, 367-387 (1948).}

\bibitem{Feynman2010}
Feynman, Richard P., Albert R. Hibbs, and Daniel F. Styer. Quantum mechanics and path integrals. Courier Corporation, 2010.

\bibitem{Kleinert1997}
Kleinert, Hagen. Quantum equivalence principle. Functional Integration. Springer, Boston, MA, 1997. 67-92.


\bibitem{Justin2010-PI}
Zinn-Justin, Jean. Path integrals in quantum mechanics. Oxford University Press, 2010.

\bibitem{Zee2010}
Zee, Anthony. Quantum field theory in a nutshell. Vol. 7. Princeton university press, 2010.


\bibitem{Feynman2000}
Feynman, Richard Phillips, and F. L. Vernon Jr. The theory of a general quantum system interacting with a linear dissipative system. 
\href{https://doi.org/10.1006/aphy.2000.6017}
{Annals of physics \textbf{281}, 547-607(2000).}

\bibitem{Weiss2012}
Weiss, Ulrich. Quantum dissipative systems. Vol. 13. World scientific, 2012.


\bibitem{Hawking-1979-QG}
Hawking, Stephen W.  The path-integral approach to quantum gravity.  General relativity. 1979.


\bibitem{Wio2013}
Wio, Horacio S. Path integrals for stochastic processes: An introduction. World Scientific, 2013.

\bibitem{Chernyak2006} 
Chernyak, Vladimir Y., Michael Chertkov, and Christopher Jarzynski.  Path-integral analysis of fluctuation theorems for general Langevin processes.  
\href{https://doi.org/10.1088/1742-5468/2006/08/p08001}
{Journal of Statistical Mechanics: Theory and Experiment \textbf{2006}, P08001 (2006).	}

\bibitem{Gennes1979}
De Gennes, Pierre-Gilles, and Pierre-Gilles Gennes. Scaling concepts in polymer physics. Cornell university press, 1979.

\bibitem{Kleinert2009}
Kleinert, Hagen. Path integrals in quantum mechanics, statistics, polymer physics, and financial markets. World scientific, 2009.

\bibitem{Linetsky1997}
Linetsky, Vadim.  The path integral approach to financial modeling and options pricing.  
\href{https://doi.org/10.1023/A:1008658226761}
{Computational Economics \textbf{11}, 129-163(1997).}

\bibitem{wiener1923differential}
Wiener, Norbert.  Differential-Space.  
\href{https://doi.org/10.1002/sapm192321131}
{Journal of Mathematics and Physics \textbf{2}, 131-174(1923).}

\bibitem{Kac1951}
Kac, Mark. On some connections between probability theory and differential and integral equations.
\href{https://digitalassets.lib.berkeley.edu/math/ucb/text/math_s2_article-15.pdf}
{ Proceedings of the second Berkeley symposium on mathematical statistics and probability. University of California Press, 1951.}



%



\bibitem{OM1953-1} 
Onsager, Lars, and Stefan Machlup.  Fluctuations and irreversible processes.  
\href{https://doi.org/10.1103/physrev.91.1505}
{Physical Review \textbf{91}, 1505(1953).}

\bibitem{MO1953-2} 
Machlup, Stefan, and Lars Onsager.  Fluctuations and irreversible process. II. Systems with kinetic energy.  
\href{https://doi.org/10.1103/physrev.91.1512}
{Physical Review \textbf{91}, 1512(1953).}

\bibitem{Glimm-Jaffe-book-2012}
Glimm, James, and Arthur Jaffe. Quantum physics: a functional integral point of view. 
\href{https://doi.org/10.1063/1.2914804}
{Physics Today \textbf{35}, 10, 82 (1982)}

\bibitem{Dewitt1957}
DeWitt, Bryce S.  Dynamical theory in curved spaces. I. A review of the classical and quantum action principles.  
\href{https://doi.org/10.1103/revmodphys.29.377}
{Reviews of modern physics \textbf{29}, 377(1957).}

\bibitem{Bastianelli2006}
Bastianelli, Fiorenzo, and Peter Van Nieuwenhuizen. Path integrals and anomalies in curved space. Cambridge University Press, 2006.

\bibitem{Salomonson1977}
Salomonson, Per.  When does a non-linear point transformation generate an extra O ($\hbar^2$) potential in the effective Lagrangian?.  
\href{https://doi.org/10.1016/0550-3213(77)90165-1}
{Nuclear Physics B \textbf{121},  433-444(1977).}

\bibitem{Gervais1976}
Gervais, J-L., and A. Jevicki.  Point canonical transformations in the path integral.  
\href{https://doi.org/10.1016/0550-3213(76)90422-3}
{Nuclear Physics B \textbf{110}  93-112(1976).}

\bibitem{Apfeldorf1996}
Apfeldorf, Karyn M., and Carlos Ordonez.  Coordinate redefinition invariance and `extra' terms.  
\href{https://doi.org/10.1016/0550-3213(96)00451-8}
{Nuclear Physics B \textbf{479},515-526(1996).}

\bibitem{Karki1997}
K\"arki, Topi and Niemi, Antti J. Supersymmetric quantum mechanics and the DeWitt effective action.
\href{https://doi.org/10.1103/PhysRevD.56.2080}
{Phys. Rev. D \textbf{56},2080--2085 (1997)}

\bibitem{Boer1996}
de Boer, J., Peeters, B., Skenderis, K., \& van Nieuwenhuizen, P. Loop calculations in quantum mechanical non-linear sigma models with fermions and applications to anomalies. 
\href{https://doi.org/10.1016/0550-3213(95)00593-5}
{Nuclear Physics B \textbf{459},  631-692(1996).}

\bibitem{Langouche-book}
Langouche, Flor, Dirk Roekaerts, and Enrique Tirapegui. Functional integration and semiclassical expansions. Vol. 10. Springer Science \& Business Media, 2013.

\bibitem{Ding2020} 
Mingnan Ding, Zhanchun Tu, and Xiangjun Xing.  Covariant formulation of nonlinear Langevin theory with multiplicative Gaussian white noises.  
\href{https://doi.org/10.1103/PhysRevResearch.2.033381}
{Physical Review Research \textbf{2}, 033381(2020).}

\bibitem{Ding2021}
Mingnan Ding and Xiangjun Xing. Covariant Non-equilibrium Thermodynamics for Small Systems. 
\href{https://doi.org/10.48550/arXiv.2105.14534}
{arXiv:2105.14534 [cond-mat.stat-mech]}


\bibitem{Hanggi-1980}
H\"anggi, P.  Connection between deterministic and stochastic descriptions of non-linear systems.  
\href{http://www.physik.uni-augsburg.de/theo1/hanggi/Papers/29.pdf}
{ {Physica} Acta \textbf{53}, 491-496 (1980).}


\bibitem{Sasa2017} 
Itami, Masato, and Shin-ichi Sasa.  Universal form of stochastic evolution for slow variables in equilibrium systems.
\href{https://doi.org/10.1007/s10955-017-1738-6}
{Journal of Statistical Physics \textbf{167}, 46-63(2017).}


\bibitem{Lau2007}
Lau, Andy WC, and Tom C. Lubensky.  State-dependent diffusion: Thermodynamic consistency and its path integral formulation.  
\href{https://doi.org/10.1103/PhysRevE.76.011123}
{Physical Review E \textbf{76}, 011123(2007).}


\bibitem{van-Kampen-stochastic}
Van Kampen, Nicolaas Godfried. {Stochastic Processes in Physics and Chemistry, 3rd ed.,} Elsevier, {2007}.


\bibitem{Edwards1964} 
Edwards, Samuel Frederick, and Y. V. Gulyaev.  Path integrals in polar co-ordinates.  
\href{https://doi.org/10.1098/rspa.1964.0100}
{Proceedings of the Royal Society of London. Series A. Mathematical and Physical Sciences \textbf{279}, 229-235(1964).}

%





\bibitem{McLaughlin1971}
D. W. McLaughlin and L. S. Schulman. Path Integrals in Curved Spaces. 
\href{https://doi.org/10.1063/1.1665567}
{Journal of Mathematical Physics \textbf{12}, 2520 (1971).}

\bibitem{Grosche1987}
C. Grosche and F. Steiner. Path integrals on curved manifolds. 
\href{https://doi.org/10.1007/bf01630607}
{Zeitschrift f{\"u}r Physik C Particles and Fields 36, 699 (1987).}

\bibitem{Hanggi1989}
P. H{\"a}nggi. Path integral solutions for non-Markovian processes. 
\href{https://doi.org/10.1007/bf01308011}
{Zeitschrift f{\"u}r Physik B Condensed Matter \textbf{75}, 275 (1989).}

\bibitem{Dekker1976}
H. Dekker. On the functional integral for generalized Wiener processes and nonequilibrium phenomena.
\href{https://doi.org/10.1016/0378-4371(76)90028-5}
{Physica A: Statistical Mechanics and its Applications \textbf{85}, 598 (1976).}

\bibitem{Fox1986}
R. F. Fox. Functional-calculus approach to stochastic differential equations. 
\href{https://doi.org/10.1103/physreva.33.467}
{Phys. Rev. A \textbf{33}, 467 (1986).}

\bibitem{Durr1978}
D. D{\"u}rr and A. Bach. The Onsager-Machlup function as Lagrangian for the most probable path of a diffusion process. 
\href{https://doi.org/10.1007/bf01609446}
{Commun.Math. Phys. \textbf{60}, 153 (1978).}

\bibitem{Alfaro1990}
J. Alfaro and P. H. Damgaard. Field transformations, collective coordinates and BRST invariance. 
\href{https://doi.org/10.1016/0003-4916(90)90230-l}
{Annals of Physics \textbf{202}, 398 (1990).}


\bibitem{Inoue1985}
A. Inoue and Y. Maeda. On integral transformations associated with a certain Lagrangian-as a prototype of quantization. 
\href{https://doi.org/10.2969/jmsj/03720219}
{J. Math. Soc. Japan \textbf{37}, 219 (1985).}

\bibitem{Grosche1994}
Grosche, Christian.  Path integration and separation of variables in spaces of constant curvature in two and three dimensions. 
\href{https://doi.org/10.1002/prop.2190420602}
{Fortschritte der Physik/Progress of Physics \textbf{42}, 509-584(1994).}

\bibitem{Lecheheb2007}
Lecheheb, A., M. Merad, and T. Boudjedaa.  Path integral treatment for a Coulomb system constrained on D-dimensional sphere and hyperboloid.  
\href{https://doi.org/10.1016/j.aop.2006.08.003}
{Annals of Physics \textbf{322}, 1233-1246(2007).}

\bibitem{Tisza1957}
Tisza, Laszlo, and Irwin Manning.  Fluctuations and irreversible thermodynamics.  
\href{https://doi.org/10.1103/physrev.105.1695}
{Physical Review \textbf{105}, 1695(1957).}

\bibitem{Bhagwat1993}
Bhagwat, K. V., Dinkar C. Khandekar, and Shilpa V. Lawande. Path integral methods and their applications. World Scientific, 1993.

\bibitem{Dekker1979}
Dekker, H. Functional integration and the Onsager-Machlup Lagrangian for continuous Markov processes in Riemannian geometries. 
\href{https://doi.org/10.1103/physreva.19.2102}
{Physical Review A \textbf{19}, 2102(1979).}

\bibitem{Janssen1976}
Janssen, Hans-Karl.  On a Lagrangean for classical field dynamics and renormalization group calculations of dynamical critical properties.  
\href{https://doi.org/10.1007/bf01316547}
{Zeitschrift f{\"u}r Physik B Condensed Matter \textbf{23}, 377-380(1976).}

\bibitem{Adib2008}
Adib, Artur B.  Stochastic actions for diffusive dynamics: Reweighting, sampling, and minimization. 
\href{https://doi.org/10.1021/jp0751458}
{ The Journal of Physical Chemistry B \textbf{112}, 5910-5916 (2008).}

\bibitem{Arnold2000}
Arnold, Peter.  Symmetric path integrals for stochastic equations with multiplicative noise. 
\href{https://doi.org/10.1103/physreve.61.6099}
{Physical Review E \textbf{61},  6099(2000).}

\bibitem{Calisto2002} 
Calisto, H., and E. Tirapegui.  Comment on `Symmetric path integrals for stochastic equations with multiplicative noise'.  
\href{https://doi.org/10.1103/PhysRevE.65.038101}
{ Physical Review E \textbf{65}, 038101(2002).}


\bibitem{Langouche1978} 
Langouche, F., D. Roekaerts, and E. Tirapegui.  On the path integral solution of the master equation.
\href{https://doi.org/10.1016/0375-9601(78)90614-x}
{Physics Letters A \textbf{68},  418-420(1978).}

\bibitem{Langouche1978-1}
Langouche, F., D. Roekaerts, and E. Tirapegui. Functional integral methods for random fields. Stochastic Processes in Nonequilibrium Systems. Springer, Berlin, Heidelberg, 1978. 316-329.


\bibitem{Tang2014}
Tang, Ying, Ruoshi Yuan, and Ping Ao.  Summing over trajectories of stochastic dynamics with multiplicative noise.  
\href{https://doi.org/10.1063/1.4890968}
{The Journal of chemical physics \textbf{141}, 044125(2014).}



\bibitem{Cugliandolo2017} 
Cugliandolo, Leticia F., and Vivien Lecomte.  Rules of calculus in the path integral representation of white noise Langevin equations: the Onsager-Machlup approach.  
\href{https://doi.org/10.1088/1751-8121/aa7dd6}
{Journal of Physics A: Mathematical and Theoretical \textbf{50} , 345001(2017).}

\bibitem{Cugliandolo2018}
Cugliandolo, Leticia F., Vivien Lecomte, and Fr\'ed\'eric Van Wijland. Building a path-integral calculus. 
\href{https://hal.archives-ouvertes.fr/hal-01823989/document}
{2018. hal-01823989}

\bibitem{Cugliandolo2019} 
Cugliandolo, Leticia F., Vivien Lecomte, and Fr\'ed\'eric Van Wijland.  Building a path-integral calculus: a covariant discretization approach. 
\href{https://doi.org/10.1088/1751-8121/ab3ad5}
{Journal of Physics A: Mathematical and Theoretical \textbf{52}, 50LT01(2019).}

\bibitem{Haken1976}
Haken, H.  Generalized Onsager-Machlup function and classes of path integral solutions of the Fokker-Planck equation and the master equation. 
\href{https://doi.org/10.1007/bf01360904} 
{Zeitschrift fur Physik B Condensed Matter \textbf{24}, 321-326(1976).}

\bibitem{Wissel1979} 
Wissel, C.  Manifolds of equivalent path integral solutions of the Fokker-Planck equation.
\href{https://doi.org/10.1007/bf01321245}  
{Zeitschrift f{\"u}r Physik B Condensed Matter \textbf{35}, 185-191(1979).}


\bibitem{Deininghaus1979} 
Deininghaus, U., and R. Graham.  Nonlinear point transformations and covariant interpretation of path integrals.
\href{https://doi.org/10.1007/bf01322143}
{Zeitschrift f{\"u}r Physik B Condensed Matter \textbf{34}, 211-219(1979).}

\bibitem{Janssen1992}
Janssen, H. K. On the renormalized field theory of nonlinear critical relaxation.
{From Phase Transitions To Chaos: Topics in Modern Statistical Physics. 1992. 68-91.}

\bibitem{Gardiner-book}
Gardiner, Crispin W. Handbook of stochastic methods. Vol. 3. Berlin: springer, 1985.

\bibitem{Jacobs2010-book}
Jacobs, Kurt. Stochastic processes for physicists: understanding noisy systems. Cambridge University Press, 2010.

\bibitem{Weiss(1978)}
Weiss, U.  Operator ordering schemes and covariant path integrals of quantum and stochastic processes in curved space.
\href{https://doi.org/10.1007/bf01321096}
{  Zeitschrift f{\"u}r Physik B Condensed Matter \textbf{30}, 429-436(1978).}

\bibitem{Graham1977} 
Graham, Robert.  Path integral formulation of general diffusion processes. 
\href{https://doi.org/10.1007/bf01312935}
{ Zeitschrift f{\"u}r Physik B Condensed Matter \textbf{26}, 281-290(1977).}

\bibitem{Graham(19772)} 
Graham, Robert.  Covariant formulation of non-equilibrium statistical thermodynamics.  
\href{https://doi.org/10.1007/bf01570750}
{Zeitschrift f{\"u}r Physik B Condensed Matter \textbf{26}, 397-405(1977).}


\bibitem{Kerler1978}
Kerler, W.  Definition of path integrals and rules for non-linear transformations.
\href{https://doi.org/10.1016/0550-3213(78)90193-1 } 
{ Nuclear Physics B \textbf{139}, 312-326(1978).}

\bibitem{Hirshfeld-1978}
Hirshfeld, Allen C. Canonical and covariant path integrals.
\href{https://doi.org/10.1016/0375-9601(78)90550-9 }
{Physics Letters A \textbf{67}, 5-8(1978).}



%
%
%
%
%









%
%
%







\end{thebibliography}

\onecolumn\newpage
\appendix
\vspace{-5mm}

\section{Proof of Lemma 3}
\label{transitionformula}
We want to prove that the following family of  time-slice Green's function (TSGF) is equivalent to each other for all $\alpha \in [0,1]$, in the sense of Lemma 1:
\ba
{\mathcal G}^{\alpha} (\xv | \xv_0; \Delta t) d\mu(\xv) &=& 
 \frac{d\mu(\xv) }{  \sqrt{ (4\pi \Delta t)^d D(\xv_{\alpha} )}} \exp \bigg\{
 \nonumber \\ 
 &-& \Big( \Delta x^i - F^i(\xv_{\alpha})\Delta t 
+ 2 \alpha \partial_l D^{il}(\xv_{\alpha}) \Delta t \Big)
 \frac{D_{ij}^{-1}(\xv_{\alpha})}{4\Delta t}
  \Big( \Delta x^j - F^j(\xv_{\alpha})\Delta t 
+ 2 \alpha \partial_k D^{jk}(\xv_{\alpha}) \Delta t \Big)
\nonumber \\ 
&- & \alpha \, \partial_i F^i( \xv_{\alpha}) \Delta t 
+ \alpha^2 \partial_i \partial_j D^{ij}(\xv_{\alpha} )\Delta t
- \Phi (\xv_{\alpha}) \Delta t \,\,\, \bigg\},
\label{Wissel}
\ea
with $d\mu(\xv) = d^dx$ and $\xv_{\alpha} = \xv+ \alpha \Delta \xv_0$. Here and below we use $D(\xv)$ to denote the determinant of the symmetric matrix $D^{ij}(\xv)$.
We only need to prove that Eq.~(\ref{Wissel}) is equivalent to the $\alpha = 0$ version: 
\ba
{\mathcal G}^0 (\xv | \xv_0; \Delta t) d\mu(\xv)  &=& 
 \frac{d\mu(\xv)    } {  \sqrt{ (4\pi \Delta t)^d D(\xv_0 )}}
 \exp \bigg\{ 
-  \Big( \Delta x^i - F^i(\xv_0,t)\Delta t  \Big)
 \frac{D_{ij}^{-1}(\xv_0)}{4\Delta t}
\Big( \Delta x^j - F^j(\xv_0,t)\Delta t  \Big)
-  \Phi(\xv_0) \Delta t \bigg\}, 
\nonumber\\
\label{Wissel-alpha-0}
\ea
which was already shown in Eq.~(\ref{Wissel-alpha-0-main-1}).  
These TSGFs are understood as functions of $\xv$ with $\xv_0, \Delta t$ serving as parameters.  

According to Lemma 1, we only need to show that Eqs.~(\ref{Wissel-alpha-0}) and (\ref{Wissel}) have equal moments up to the order $ \Delta t$.  Since  $\Delta \xv \sim \sqrt{\Delta t}$, we see that the difference between $\Phi(\xv_0) \Delta t$ and $\Phi(\xv_{\alpha}) \Delta t$ is of order $ \Delta t^{3/2}$ and hence makes no contribution.  It then follows that we only need to prove the equivalence for the case $\Phi = 0$. Below we will set $\Phi = 0$.  Consequently the zeroth order moments of ${\mathcal G}^{\alpha} (\xv | \xv_0; \Delta t) d\mu(\xv)$ equals unity, and the TSGFs can be understood as a classical probability distributions  up to the order $\Delta t$.

It is  convenient to introduce a set of scaled variables $\xi^i$ via 
\ba
\Delta x^i =   \xi^i  \sqrt{ \Delta t }  + F^i(\xv_0) \Delta t.
\label{xi-def-1}
\ea
Equation (\ref{Wissel-alpha-0}) can then be rewritten as (recall we have set $\Phi = 0$)
\ba
p_0({\boldsymbol \xi}; \xv_0) d^d \xi = 
\frac{ d^d \xi}{ \sqrt{(4 \pi)^{d}  D(\xv_0)} }
e^{- \xi^i  D_{ij}^{-1}(\xv_0 ) \xi^j/4 },
\label{p-alpha-0-1} 
\ea
with shows that ${\boldsymbol \xi}$ have vanishing  averages and variances $2 D^{ij}(\xv)$.   We can similarly rewrite Eq.~(\ref{Wissel}) in terms of ${\boldsymbol \xi}, \xv_0$ and $\Delta t$ in the following form: 
\ba
p_{\alpha}({\boldsymbol \xi}; \Delta t, \xv_0) d^d \xi &=& 
\frac{ d^d \xi}{ \sqrt{(4 \pi)^{d} \det D_{ij}(\xv_0) } }
e^{- \xi^i  D_{ij}^{-1}(\xv_0 ) \xi^j/4 -  \delta \mathcal A_\alpha (\boldsymbol \xi ; \Delta t, \xv_0) },
\label{p-alpha-def}
\label{A-p-alpha-def} 
\ea
where $ \delta \mathcal A_\alpha (\boldsymbol \xi ; \Delta t, \xv_0)$ can be expanded in terms of $\Delta t$, up the first order.  Higher order terms do not contribute to the continuum limit according to Lemma 1.   The result is 
\begin{subequations}
 \label{A-split-all} 
 \ba
\delta \mathcal A_\alpha  &=&  
 \mathcal A^{{1}/{2} }  \sqrt{\Delta t}
 +  \mathcal A^{1}  \Delta t
 + \cdots,\\
 \mathcal A^{1/2 }   &=&
- \frac{\alpha}{ 4 } \left(  \Dm^{-1}\
 (\partial_k \Dm) \Dm^{-1}\right)_{ij} \xi ^i \xi ^j \xi ^k
+   \frac{\alpha}{2} \mbox{Tr}
 \big(  \Dm^{-1} \partial_i \Dm \big) \xi ^i 
 +\alpha \,  D^{-1}_{ij}  (\partial_l D^{jl})  \xi^i ,
 \label{A-1/2}  \\
  \mathcal A^{1} &=&  \mathcal A^{1,1} + \mathcal A^{1,2} 
  + \mathcal A^{1,3} + \mathcal A^{1,4} 
  + {\alpha} \, \partial_i F^i- \alpha^2 \partial_i \partial_j D^{ij} ,
   \label{A1split}  \\
   \mathcal A^{1, 1}  &=& - \frac{\alpha}{4}  \left( \Dm^{-1} (\partial_k  \Dm) 
  \Dm^{-1} \right)_{ij} F^k   \xi^i \xi^j 
+  \frac{\alpha}{2} \mbox{Tr} \big(  \Dm^{-1} (\partial_i \Dm)  \big)F^i ,
 \label{A-11} \\
\mathcal A^{1,2}   &=&  - \frac{ \alpha }{2}  (\partial_k F^i)   D^{-1}_{ij} \xi^k \xi^j , 
 \label{A-12}  \\
 \mathcal A^{1, 3}  &=& -\frac{ \alpha^2}{8 }
  \left( \Dm^{-1} ( \partial_i \partial_j \Dm  ) \Dm^{-1} \right)_{lm} \xi^i \xi^j \xi^l \xi^m 
  + \alpha^2 D^{-1}_{ij} ( \partial_k \partial_l D^{ik})\xi^j\xi^l
+ \frac{\alpha^2}{4} \mbox{Tr}  \big( 
 \Dm^{-1} \partial_i \partial_j \Dm \big)  \xi^i\xi^j ,
 \label{A-13} \\
 \mathcal A^{1, 4} &=&
 \frac{\alpha^2}{4} ( \Dm^{-1} (\partial_i \Dm) \Dm^{-1} 
 (\partial_j \Dm) \Dm^{-1})_{mn} \xi^i\xi^j\xi^m\xi^n
 -  \frac{\alpha^2}{4} \mbox{Tr} \left( \Dm^{-1} (\partial_i \Dm) 
\Dm^{-1} (\partial_j \Dm) \right) \xi^i\xi^j 
\nonumber \\ 
&  - &  \alpha^2 (  \partial_l D^{il}) 
\left(  \Dm^{-1} (\partial_k \Dm) \Dm^{-1} \right)_{ij} \xi^k \xi^j
+  \alpha^2 (  \partial_l D^{il} ) D^{-1}_{ ij} (  \partial_m D^{jm}),
 \label{A-14} 
\ea
\end{subequations}
where it is understood that all functions in Eqs.~(\ref{A-split-all}) are evaluated at $\xv_0$.  Note that $\delta \mathcal A_\alpha $ vanishes as either $\alpha$ or $\Delta t$ goes to zero. 

Expanding Eq.~(\ref{A-p-alpha-def}) in terms of $\Delta t$, we find 
\ba
 p_{\alpha} = p_0 \left( 1-  \mathcal A^{1/2} \sqrt{\Delta t}
 +  \left[ \frac{1}{2} \left( \mathcal A^{ {1}/{2}  } \right)^2
-  \mathcal A^{1} \right] {\Delta t} 
+ O(\Delta t^{3/2} )   \right). 
\ea
Using this and Eq.~(\ref{p-alpha-0-1}), we can expand the moments of $ p_{\alpha} $ in terms of $\Delta$:
\begin{subequations}
\label{moments-p_alpha}
\ba
\int p_{\alpha} d^d\xi &=&  1 +
 \left\langle \frac{1}{2} \left( \mathcal A^{ {1}/{2}  } \right)^2
-  \mathcal A^{1} \right\rangle_0  \Delta t 
+ o(\Delta t), \\
\int p_{\alpha} \xi^i d^d\xi &=& 
-  \left\langle  \mathcal A^{ {1}/{2}}  \xi^i \right\rangle_0  \Delta t^{1/2}
+  O(\Delta t), \\
\int p_{\alpha} \xi^i \xi^j d^d\xi &=& 2 D^{ij}  
+  \left\langle  \mathcal A^{ {1}/{2}}  \xi^i \xi^j 
\right\rangle_0  \Delta t^{1/2} +  O(\Delta t),
\ea
\end{subequations}
where $\langle \,  \cdot \, \rangle$ means average over $p_0({\boldsymbol \xi}; \xv_0)$ as defined in Eq.~(\ref{p-alpha-0-1}):
\ba
\langle f({\boldsymbol \xi}) \rangle_0 \equiv 
 \int  p_0 ({\boldsymbol \xi}; \xv)   f({\boldsymbol \xi}) d^d \xi. 
\ea
  It is evident that all moments of $ p_{\alpha} $ can be expanded in terms of $\Delta t$.  
  
  On the other hand, using Eq.~(\ref{xi-def-1}), we can relate the moments of ${\mathcal G}_{\alpha} $ to those of $p_{\alpha}$.  Combining with Eqs.~(\ref{moments-p_alpha}) we can expand the moments of Eq.~(\ref{Wissel}) as
\begin{subequations}
\label{conditions-moments}
\ba
\int  {\mathcal G}^{\alpha} d \mu(\xv)  &=& 
 1 +  \left\langle \frac{1}{2} \left( \mathcal A^{ {1}/{2}  } \right)^2
-  \mathcal A^{1} \right\rangle_0  \Delta t 
+ o(\Delta t), 
\label{conditions-moments-0}\\
\int   {\mathcal G}^{\alpha} \Delta x^i  d \mu(\xv)  &=&
F^i(\xv_0) \Delta t  
-  \left\langle  \mathcal A^{ {1}/{2}}  \xi^i \right\rangle_0  \Delta t 
+ o( \Delta t), 
\label{conditions-moments-1}\\
\int {\mathcal G}^{\alpha} \Delta x^i \Delta x^j  d \mu(\xv)  &=& 
2 D^{ij}(\xv_0) \Delta t +  o( \Delta t), 
\label{conditions-moments-2}\\
\int {\mathcal G}^{\alpha} \Delta x^i \Delta x^j \Delta x^k \cdots   d \mu(\xv)   &=& o( \Delta t). 
\ea
\end{subequations}
Recall that we  need to prove that, up to $\Delta t$, these moments equal to those of ${\mathcal G}^0 (\xv | \xv_0; \Delta t) d\mu(\xv)$, which are already shown in Eqs.~(\ref{three-moments-0}) (with $\Phi$ set to zero, again).   By inspection, we see that equality of zero-th and first order moments are guaranteed as long as we can prove the following identities:
\ba
\left\langle \mathcal A^{1} \right\rangle_0 
 - \frac{1}{2} \left\langle \left( \mathcal A^{ {1}/{2}  } \right)^2
\right\rangle_0 
&=& 0, 
\label{conditions-moments-0-3} \\
  \left\langle  \mathcal A^{ {1}/{2}}  \xi^i \right\rangle_0 
  &=& 0.
  \label{ave-A-1/2}
\ea
 Equality of second moments and higher moments are automatically guaranteed.

Proof of Eq.~(\ref{conditions-moments-0-3}) is very complicated, and is presented in Sec.~\ref{sec:proof-normalization}.   Here let us prove Eq.~(\ref{ave-A-1/2}). Since $p_0$ is Gaussian in $\xi$, the LHS of Eq.~(\ref{ave-A-1/2})  can be calculated using Wick's theorem.  Substituting  Eq.~(\ref{A-1/2}) into Eq.~(\ref{ave-A-1/2}) and using Wick's theorem, we find
\ba
  \left\langle  \mathcal A^{ {1}/{2}}  \xi^i \right\rangle_0 
&=& - \frac{\alpha}{ 4 } \left(  \Dm^{-1} 
 (\partial_k \Dm) \Dm^{-1}\right)_{lj} 
\langle \xi ^i \xi ^j \xi ^k \xi^l \rangle 
+   \frac{\alpha}{2} \mbox{Tr}
 \big(  \Dm^{-1} \partial_l \Dm \big)
   \langle \xi^i \xi ^l \rangle 
 + \alpha \,  D^{-1}_{kj}  (\partial_l D^{jl})
 \langle  \xi^i \xi^k \rangle  
 \nonumber\\
 &=& - \alpha \left( \Dm^{-1} 
(\partial_k \Dm ) \Dm^{-1}\right)_{lj}
 ( D^{ij} D^{kl} + D^{ik} D^{jl} + D^{il} D^{jk} )
 + \alpha D^{il} D^{-1}_{jk} \partial_l D^{jk} 
 + 2 \alpha D^{-1}_{kj}  (\partial_l D^{jl}) D^{ik}
\nonumber \\
& = & - \alpha ( \partial_k D^{ki} + D^{ik} D^{-1}_{lm}\partial_k D^{ml}  + \partial_k D^{ik} )
+\alpha D^{il} D^{-1}_{jk} \partial_l D^{jk} + 2 \alpha \partial_l D^{il}  = 0,
\ea
which establishes the validity of Eq.~(\ref{ave-A-1/2}). 

%

\vspace{-3mm}
\subsection{Proof of Eq.~(\ref{conditions-moments-0-3})}
\label{sec:proof-normalization}
Using Eq.~(\ref{A1split}) we may rewrite the LHS of Eq.~(\ref{conditions-moments-0-3}) as
\ba
\left\langle \mathcal A^{1} \right\rangle_0 
 - \frac{1}{2} \left\langle \left( \mathcal A^{ {1}/{2}  } \right)^2
\right\rangle_0 \!\! &=&  
   \left\langle  \mathcal A^{1,1} \right\rangle_0 
  + \left\langle \mathcal A^{1,2} \right\rangle_0 
  + \left\langle \mathcal A^{1,3} \right\rangle_0 
  + {\alpha} \, \partial_i F^i 
  - \alpha^2 \partial_i \partial_j D^{ij} 
   + \left\langle \mathcal A^{1,4} \right\rangle_0 
     - \frac{1}{2} \left\langle \left( \mathcal A^{ {1}/{2}  } \right)^2 \right\rangle_0 ,
     \quad\quad
   \label{A1split-ave} 
\ea
Let us calculate these averages using Eqs.~(\ref{A-split-all}) and Wick's theorem:
\ba
\langle  \mathcal A^{1,1}  \rangle_0 &=& 
 - \frac{1}{4}
 \left(\alpha \Dm^{-1} (\partial_k  \Dm)  \Dm^{-1} \right)_{ij} 
 F_1^k  \langle \xi^i \xi^j  \rangle_0
 + \frac{1}{2} \mbox{Tr} \big( \alpha \Dm^{-1} (\partial_i \Dm) \big)F^i
\nonumber \\ 
  &=& - \frac{\alpha}{2} D^{ij} D^{-1}_{im} (\partial_k D^{ml}) D^{-1}_{lj} F^k
 + \frac{\alpha}{2} D_{jk}^{-1}( \partial_i D^{jk} )F^i =0, \\
\left \langle \mathcal A^{1,2}  \right \rangle_0
  &=& - \frac{ \alpha }{2}  (\partial_k F^i)
    D^{-1}_{ij}\langle \xi^k \xi^j  \rangle_0
 = - \frac{ \alpha}{2} \, 2 D^{kj }D^{-1}_{ij}\partial_k F^i 
 = -\alpha\, \partial_i F^i, \\
\left\langle \mathcal A^{1,3} \right\rangle_0
 &=& -\frac{ \alpha^2}{8 } D_{lm}^{-1} (\partial_i \partial_j D^{mn}) 
 \langle \xi^i \xi^j \xi^l \xi^n \rangle_0
 + \alpha^2 D^{-1}_{ij} ( \partial_k \partial_l D^{ik}) \langle \xi^j\xi^l\rangle_0
 + \frac{\alpha^2}{4} \mbox{Tr}  \big( 
 \Dm^{-1} (\partial_i \partial_j \Dm) \Dm^{-1}\big)
  \langle  \xi^i\xi^j  \rangle_0
\nonumber\\ 
& = & -\frac{ \alpha^2}{2} D_{lm}^{-1}( \partial_i \partial_j D^{mn} )
\big(D^{ij} D^{ln} + D^{il} D^{jn} 
+ D^{in } D^{jl}\big) 
 + 2 \alpha^2 D^{-1}_{ij} ( \partial_k \partial_l D^{ik}) D^{jl}
+\frac{\alpha^2}{2}  D^{ij} D^{-1}_{kl}\partial_i\partial_j D^{lk}
 \nonumber\\ 
 &=& \alpha^2 \partial_i \partial_j D^{ij},
\ea
Summing up these three equations, we find that the sum of the first five terms in Eq.~(\ref{A1split-ave}) vanishes identically:
\ba
  \left\langle  \mathcal A^{1,1} \right\rangle_0 
  + \left\langle \mathcal A^{1,2} \right\rangle_0 
  + \left\langle \mathcal A^{1,3} \right\rangle_0 
  + {\alpha} \, \partial_i F^i 
  - \alpha^2 \partial_i \partial_j D^{ij} = 0. 
 \label{sum-5-terms}
\ea 
Additionally we have
\ba
\langle  \mathcal A^{1,4}\rangle_0&=&
 \frac{\alpha^2}{4} ( \Dm^{-1} (\partial_i \Dm) \Dm^{-1} 
 (\partial_j \Dm) \Dm^{-1})_{mn}  \langle \xi^i\xi^j\xi^m\xi^n \rangle_0
 -  \frac{\alpha^2}{4} \mbox{Tr} \left( \Dm^{-1} (\partial_i \Dm) 
\Dm^{-1} (\partial_j \Dm) \right) \langle \xi^i\xi^j \rangle_0
\nonumber \\ 
&   & -  \alpha^2 (  \partial_l D^{il}) 
\left(  \Dm^{-1} (\partial_k \Dm) \Dm^{-1} \right)_{ij}  \langle \xi^k \xi^j \rangle_0
+  \alpha^2 (  \partial_l D^{il} ) D^{-1}_{ ij} (  \partial_m D^{jm})
\nonumber \\
& = & \alpha^2 \bigg[
 (\partial_m D^{jk}) D^{-1} _{kl} (\partial_n D^{lr} )
  D^{-1}_{rj} D^{mn}
 +( \partial_m D^{mk} )D^{-1} _{kl}( \partial_n D^{ln}) 
  + ( \partial_m D^{sk} )D^{-1} _{kl} (\partial_s D^{lm})
\nonumber  \\
&& -\frac{1}{2} D^{-1} _{ik}( \partial_s D^{kl}) 
D^{-1}_{lm}( \partial_t D^{mi } )D^{st}
 -2( \partial_s D^{st} )D^{-1} _{tj}( \partial_l D^{jl})
 +( \partial_l D^{il} )D^{-1} _{ij} (\partial_k D^{jk})
 \bigg]
\nonumber \\
& = & \alpha^2 \bigg[ 
\frac{1}{2}(\partial_m D^{jk}) D^{-1} _{kl}( \partial_n D^{lr} )D^{-1}_{rj} D^{mn}
+ ( \partial_m D^{sk} )D^{-1} _{kl} (\partial_s D^{lm}) \bigg].
\label{A14result}
\ea
Similarly using Eq.~(\ref{A-14}) and Wick's theorem, we can calculate the last term in Eq.~(\ref{A1split-ave}): 
\ba
  \frac{1}{2} \left\langle\left( \mathcal A^{ 1/2}\right)^2\right\rangle_0
&=& \alpha^2 \left[ \langle  \mathcal  A^{2,1} \rangle_0
+ \langle \mathcal   A^{2,2} \rangle_0
+ \langle \mathcal  A^{2,3} \rangle_0
+ \langle  \mathcal  A^{2,4}\rangle_0\right], \\
\langle \mathcal   A^{2,1} \rangle_0& \equiv & 
\frac{1}{4} D^{ij} D^{-1} _{mn}
 (\partial_i D^{mn}) D^{-1}_{st}( \partial_j D^{st})
+ (\partial_l D^{jl} )D^{-1} _{jm} (\partial_n D^{mn})
+ D^{-1} _{kl} (\partial_m D^{kl} )(\partial_n D^{mn}),
\nonumber \\
\langle \mathcal   A^{2,2} \rangle_0& \equiv & 
\frac{1}{4} D^{-1} _{il}( \partial_k D^{lm} )D^{-1}_{mj}
D^{-1} _{sp}( \partial_r D^{pq} )D^{-1}_{qt} ( D^{ij} D^{kr} D^{st} 
+ \mbox{other 14 pairing terms}),
\label{A22}
\nonumber \\
\langle \mathcal   A^{2,3} \rangle& \equiv & 
- \frac{1}{2} D^{-1}_{lj} (\partial_i D^{jl} )
D^{-1} _{pm} (\partial_k D^{nm} )D^{-1}_{nq} 
( D^{ik} D^{pq}+D^{ip} D^{kq}+D^{iq} D^{pk}),
\nonumber \\
\langle \mathcal   A^{2,4} \rangle_0& \equiv &- D^{-1}_{ij} (\partial_l D^{lj} )
D^{-1} _{pm} (\partial_k D^{nm} )D^{-1}_{nq} 
( D^{ik} D^{pq}+D^{ip} D^{kq}+D^{iq} D^{pk}).
\ea

Adding up all these four terms (which demands great patience), we find
\ba
 \frac{1}{2}  \left\langle \left( \mathcal A^{ 1/2}\right)^2\right\rangle_0
 =  \alpha^2 \bigg[ 
\frac{1}{2}(\partial_m D^{jk}) D^{-1} _{kl}( \partial_n D^{lr} )D^{-1}_{rj} D^{mn}
+ ( \partial_m D^{sk} )D^{-1} _{kl} (\partial_s D^{lm}) \bigg]
= \left\langle  \mathcal A^{1,4} \right\rangle_0,
\label{equality-temp}
\ea
where in the last step we have used Eq.~(\ref{A14result}).  Finally combining Eq.~(\ref{equality-temp}) with Eq.~(\ref{sum-5-terms}), we find that Eq.~(\ref{A1split-ave}) vanishes identically.  This establishes the validity of Eq.~(\ref{conditions-moments-0-3}). 

This completes our proof of equivalence between Eqs.~(\ref{Wissel}) and (\ref{Wissel-alpha-0}).  

\vspace{4mm}

\end{document}